# Satellite-based estimates of decline and rebound in China's $CO_2$ emissions during COVID-19 pandemic

Bo Zheng[1,2,†], Guannan Geng[3,†], Philippe Ciais[1], Steven J. Davis[4], Randall V. Martin[5,6,7], Jun Meng[6,5], Nana Wu[2], Frederic Chevallier[1], Gregoire Broquet[1], Folkert Boersma[8,9], Ronald van der A[8], Jintai Lin[10], Dabo Guan[2], Yu Lei[11], Kebin He[3], Qiang Zhang[2,*]

[1]Laboratoire des Sciences du Climat et de l'Environnement, CEA-CNRS-UVSQ, Gif-sur-Yvette, France.

[2]Ministry of Education Key Laboratory for Earth System Modeling, Department of Earth System Science, Tsinghua University, Beijing, China.

[3]State Key Joint Laboratory of Environment Simulation and Pollution Control, School of Environment, Tsinghua University, Beijing, China.

[4]Department of Earth System Science, University of California, Irvine, Irvine, CA, USA.

[5]Department of Energy, Environmental & Chemical Engineering, Washington University in St. Louis, St. Louis, MO, USA.

[6]Department of Physics and Atmospheric Science, Dalhousie University, Halifax, NS, Canada.

[7]Harvard-Smithsonian Center for Astrophysics, Cambridge, MA, USA.

[8]Royal Netherlands Meteorological Institute (KNMI), De Bilt, the Netherlands.

[9]Environmental Sciences Group, Wageningen University, Wageningen, the Netherlands.

[10]Laboratory for Climate and Ocean-Atmosphere Studies, Department of Atmospheric and Oceanic Sciences, School of Physics, Peking University, Beijing, China.

[11]Chinese Academy of Environmental Planning, Beijing, China.

[*]Correspondence to: qiangzhang@tsinghua.edu.cn.

[†]These authors contributed equally to this work.

**Abstract**

Changes in $CO_2$ emissions during the COVID-19 pandemic have been estimated from indicators on activities like transportation and electricity generation. Here, we instead use satellite observations together with bottom-up information to track the daily dynamics of $CO_2$ emissions during the pandemic. Unlike activity data, our observation-based analysis can be independently evaluated and can provide more detailed insights into spatially-explicit changes. Specifically, we use TROPOMI observations of $NO_2$ to deduce ten-day moving averages of $NO_x$ and $CO_2$ emissions over China, differentiating emissions by sector and province. Between January and April 2020, China's $CO_2$ emissions fell by 11.5% compared to the same period in 2019, but emissions have since rebounded to pre-pandemic levels owing to the fast economic recovery in provinces where industrial activity is concentrated.

[*124 words*]

**One Sentence Summary:** Satellite reveals 10-day mean and spatially-explicit variations in China's $CO_2$ emissions during and after COVID-19 lockdown.

**Main Text**

In the first half of 2020, most countries in the world have imposed stringent policies to slow down the spread of Coronavirus disease 2019 (COVID-19), closing businesses and factories, restricting travel, and issuing stay-at-home orders. Such lockdown measures have helped to curb the spread of the virus (*1*) and meanwhile caused large reductions in global demand for fossil fuels (*2*). In turn, levels of nitrogen dioxide ($NO_2$) and other air pollutants have also fallen across the globe (*3-5*), and global carbon dioxide ($CO_2$) emissions declined by an estimated 8.6%, based on indicators of energy use between January and April of 2020 as compared to the same period in 2019 (*6*). If history is a guide, though, such reductions in air pollution and $CO_2$ emissions could be temporary: global $CO_2$ emissions have rebounded and resumed their former rates of growth after every financial crisis in the fossil fuel era (*7, 8*). Indeed, economic activities are already increasing in many places, and governments and central banks have already passed or proposed large economic stimulus packages to spur recovery (*9*) without any climate mitigation co-objective. Yet the COVID-19 pandemic has come at an important moment in the centuries-long timeline of fossil energy use, and for that reason as well as the pandemic's outsized disturbance to social and economic systems (*10*), it may mark a turning point in the world's energy and economic structure—one with lasting influence on the trajectory of global $CO_2$ emissions.

However, annual, country-level inventories of $CO_2$ emissions (*11*) are plainly insufficient to monitor variations in the sources and dynamics of $CO_2$ emissions during fast-evolving periods like the COVID-19 pandemic. Nor can such annual estimates support adaptive and agile climate and energy policies going forward, as countries and other jurisdictions seek to finetune such policies to achieve environmental goals in a rapidly evolving technological and economic context. Increasing the frequency and resolution of data on $CO_2$ emissions is thus a research priority. Unfortunately, the data streams required to conduct continuous carbon monitoring with the high temporal and spatial resolution are yet very limited. Bottom-up approaches to estimate daily $CO_2$ emissions at the country level using activity data and energy use indicators have recently emerged (*6, 12, 13*), but gaining access to reliable daily statistics of sector-specific fossil fuel consumption is a challenge, and activity proxies such as traffic congestion indices and heating degree days must be used in those recent studies to empirically analyze the relative changes of emissions. Moreover, such bottom-up estimates lack independent validation.

Near real-time observations from ground- and space-based platforms thus represent an attractive means of supplementing and validating bottom-up estimates with direct measurements (*14-18*). However, $CO_2$ concentrations are sparsely sampled in time and space, even for the carbon satellites (*14, 15*), and natural variability in ecosystem carbon fluxes and atmospheric transport prevents unambiguous detection of fossil fuel $CO_2$ emissions over time scales of weeks to months. Moreover, the current generation of carbon monitoring satellites often returns significant errors in single soundings data (*14, 15*). Satellite observations of $NO_2$, a species co-emitted with $CO_2$ during the combustion of fuels, have broader coverage than $CO_2$ observations, especially from the recently launched TROPOspheric Monitoring Instrument (TROPOMI) onboard the Copernicus Sentinel-5 Precursor satellite (*19*). Due to the relatively short lifetime in $NO_x$, the satellite is capable of detecting the short-term variability in $NO_2$ columns (*3, 5*). Satellite $NO_2$ columns have been widely used to retrieve the spatial patterns or temporal trends of $NO_x$ emissions (e.g., *20-23*). There are also attempts to infer $CO_2$ emissions from satellite-based $NO_2$ observations (*24-26*), however, such a method requires good knowledge of $CO_2$ to $NO_x$ emission ratios that are region-dependent, sector-specific, and dynamically changing (*27*).

Here, we develop a novel approach to infer a ten-day moving average of anthropogenic $CO_2$ emissions from TROPOMI $NO_2$ merged with bottom-up information. Details of our data sources and methods are provided in the *Supplementary Materials*. In summary, we first develop a preliminary bottom-up estimate of sectoral $NO_x$ and $CO_2$ emissions in 2020 based on the Multi-resolution Emission Inventory for China (MEIC) in 2019 (*28*) and near-real-time statistics and proxies in 2020. Based on the high-quality $NO_2$ column observations from TROPOMI, we then separate the meteorological effects and model the local sensitivity of $NO_2$ column to surface $NO_x$ emission changes (*29*) with the nested-grid GEOS-Chem chemical transport model (*30*), to assess $NO_x$ emission changes in 2020 compared to those in 2019. Next, we use the top-down estimates to correct the sectoral distribution in the preliminary bottom-up emissions, based on the emission differences revealed by the grid cells dominated by a single source sector. Finally, the TROPOMI-constrained $NO_x$ emissions and the spatiotemporal heterogeneity of emission sectoral information are combined with the spatially-explicit, sector-specific ratio maps of $CO_2$ to $NO_x$ emissions to infer ten-day moving average $CO_2$ emissions from specific sectors. We use this integrated satellite-based emissions monitoring approach to track $NO_x$ and $CO_2$ emissions in mainland China over the period from January to April 2020 as the coronavirus lockdown was imposed and relaxed.

**Daily Dynamics of $NO_x$ and $CO_2$ Emissions**

Figure 1 presents 10-day moving averages of $NO_x$ (Figures 1A and 1C) and $CO_2$ emissions (Figures 1B and 1D) from January to April in 2019 and 2020. The 2019 emissions are directly derived from the MEIC inventory, which is also the base for estimating the $NO_x$ and $CO_2$ emissions in 2020 constrained by TROPOMI observations. The 2019 emissions data show a reduction in the daily $NO_x$ and $CO_2$ by 30% and 38%, respectively, from 1st January to the Chinese New Year 2019 (5th February) and a rebound in emissions twenty days later when the holiday ended and people returned to work. The larger decrease in $CO_2$ than in $NO_x$ is because the major driver of emissions decrease is the reduced activities in the power and industry sectors (where emissions ratios of $CO_2$ to $NO_x$ are large), while transport emissions (with low $CO_2$ to $NO_x$ emission ratios) decreased much less because travel demand was still high during the holiday.

The 2020 daily emissions constrained by TROPOMI observations, however, show much faster decreases than those in 2019. During the period from 1st January to the Chinese New Year 2020 (25th January), $NO_x$ and $CO_2$ emissions are estimated to have dropped by 58% and 51%, respectively. The larger decrease in $NO_x$ emissions than in $CO_2$ is due to a much larger decline in the transport emissions than power and industrial emissions during the COVID-19 lockdown as we will show in the following analysis. The daily emissions of $NO_x$ and $CO_2$ in 2020 are estimated up to be 50% and 40%, respectively, lower than those in 2019, and they did not return to the pre-holiday levels until two months later.

The divergence of the daily emissions between 2020 and 2019 corresponds to the timeline of the virus control measures. The sharper emission declines in 2020 started on 20th January 2020, when the most stringent control measures were activated by the National Health Commission. The Wuhan lockdown began three days later on 23rd January 2020, which was followed by similar measures in the other Chinese cities within the next few days. These lockdown measures did not ease until about one month later when the lowest-risk regions and cities slowly reopened some of the less exposed industries and businesses. About two months after the Chinese New Year 2020, most of China's cities had lifted the control measures including Wuhan that reopened on 8th April after a 76-day lockdown. During the Wuhan lockdown period (grey shades in Figure 1), China's emissions were lower than the 2019 emissions by a cumulative of 892 kt $NO_x$ (21.9% net reduction, green shades in Figure 1) and 348 Mt $CO_2$ (16.2% net reduction, blue shades in Figure 1).

The total reduction in China's $CO_2$ emissions over January–April 2020 is equivalent to a −11.5% decrease over the corresponding period of 2019, a bit higher than the estimate of −7.8% (−3.6% to −12.9%) from (*6*). The largest emission reductions occurred in February, while the emissions rapidly rebounded in March and April (Table S1), and the $CO_2$ emissions in April 2020 is estimated +2.7% higher than that in 2019. The emission reductions over January–March 2020 are −15.6% compared to the the same period of 2019, which is higher than the the bottom-up estimate of −10.3% from (*13*). Overall, our TROPOMI-constrained emission estimates present larger emission reductions during January–April than the bottom-up estimates (*6*, *13*).

We also used the bottom-up approach (see Methods) to estimate daily $NO_x$ and $CO_2$ emissions in 2020 (dashed curves in Figures 2A and 2B), which are broadly comparable to the TROPOMI constrained inversions but still reveal large discrepancies. The differences in the $NO_x$ emissions mainly occur over the regions dominated by the emissions from transport (blue curve in Figure 2C) and industry (yellow curve in Figure 2C), while the power sector also contributes to the discrepancy in the $CO_2$ emissions (red curve in Figure 2D). The bottom-up inventory tends to overestimate industrial and transport emissions during the lockdown period, especially at the beginning of lockdown, while tends to underestimate emissions for the power sector when lockdown gradually lifted in March. Since the complete statistics of daily fuel combustion are not available for the bottom-up inventory (see Methods), which relies instead on the daily data for 12% of China's power plants and some proxies such as the daily traffic congestion indices and monthly industrial Gross Domestic Product (GDP) combined with the daily coal use in the power sector to represent the relative changes of daily emissions, which is inevitably associated with uncertainties.

**Drivers of $CO_2$ Emissions Drop and Rebound**

We decompose the difference in the ten-day moving average of $CO_2$ emissions between 2020 and 2019 into power, industry, transport, and residential sectors (Figures 3A and 3B). The industry sector was the major driver of the $CO_2$ emissions changes. During the Wuhan lockdown period (grey shades in Figure 3), the cumulative $CO_2$ emissions from the industry sector declined by 24.3% compared to those in 2019, accounting for 72% of the total reduction in $CO_2$ emissions for that period. The transport and power sectors are estimated to have experienced cumulative $CO_2$ emissions reductions by 31.1% and 5.0%, respectively, contributing 18% and 10% of the total $CO_2$ reduction. The emissions decline in the power sector was possibly driven by the lower demand by industry that consumes more than two-thirds of the total electricity in China. Residential emissions increased by 1%. The population-weighted heating degree day in 2020 winter was estimated 3% lower than that in 2019 due to the high air temperature, therefore the larger residential emissions in 2020 are mainly due to the more energy consumed when people were forced to stay at home, especially at the beginning of the lockdown period. After the lockdown measures lifted, the recovery of the industry and transport sectors rapidly pushed $CO_2$ emissions back to pre-lockdown levels.

The regional decomposition (Figure 3C and 3D) also confirms a $CO_2$ emission dynamics that was dominated by industry. China's provinces were classified into three categories distinguished by its increasing share of industrial GDP in the provincial total GDP and the length of lockdown. The provinces dominated by the industrial economy were more sensitive to the lockdown measures, with both a deeper drop and a faster recovery in emissions. The $CO_2$ emissions declined by 8.2% in the provinces where industrial GDP contributes less than 34% (median value of all the provinces) of the provincial total GDP (pink bars in Figure 3C and 3D). However, the emission decreases were more than twice higher in provinces where the share of industrial GDP is larger than 34%. The industrial provinces with a lockdown longer than 40 days (dark blue bars in Figures 3C and 3D) show comparable reductions of $CO_2$ emissions (−20.2%) than those (−21.2%) with a shorter lockdown period (light blue bars in Figure 3C and 3D).

**Response of Provincial Emissions to COVID-19**

Half of China's provinces are estimated to have reduced their cumulative $CO_2$ emissions by more than 15% from 23rd January 2020 to 7th April 2020 compared to the same period in 2019 (Figures 4A and 4B). Since the industrial sector was the major driver of this emission decline, provinces with larger shares of the industrial economy experienced more $CO_2$ emission reductions than others (Figure 4B). The provinces of Jiangsu and Anhui that have the largest industrial economies both reduced their $CO_2$ emissions by more than 30%. We also observe large $CO_2$ reductions in the Hubei Province whose capital is Wuhan and in Beijing and Shanghai, significantly higher than the other provinces with similar economic structures due to the stringent virus control measures in these provinces. It should also be noted that the $CO_2$ emissions from Guangxi Province located in the southwest of China are estimated to have increased by 5% during the lockdown period. This is because the generation of hydroelectric power from Jan to March in 2020 was 27% lower than that in 2019 due to the severe drought in this region, which has caused an increase in the generation of thermal power by 16% (http://www.stats.gov.cn/). The drought-induced increased fossil fuel use in power plants offset the emissions decrease due to lockdown.

With China's industry slowly recovering from the coronavirus, $CO_2$ emissions from most of China's eastern provinces have returned to their pre-COVID-19 levels by April 2020, which are higher than the emissions in the same period of 2019 (Figures 4C and 4D). The provinces whose economy is dominated by industry restarted emissions rapidly, mirroring their larger drop during the lockdown. The provinces that had experienced stringent lockdown did not rebound their $CO_2$ emissions significantly, such as Hubei, Hebei, and Tianjin, although these provinces were dominated by the industrial economy. These provinces stayed at the highest emergency response levels against the coronavirus for more than 90 days. The emissions from Beijing and Shanghai were 17% and 22%, respectively, lower than that in 2019 at this period, which exhibited comparable and even larger reductions in $CO_2$ emissions compared to the previous two months. The Guangxi province, still suffering from the drought in April 2020, increased $CO_2$ emissions by 29% compared to those in 2019, corresponding to the 30% increase in thermal power generation.

**Discussion**

This study presents the first-ever estimates of the ten-day moving average $CO_2$ emissions from satellite observations, derived from the near real-time and high spatial-temporal resolution $NO_2$ retrievals from TROPOMI, state-of-the-art chemical transport model GEOS-Chem, and $CO_2$ to $NO_x$ source emission ratios from a detailed inventory of Chinese emissions, MEIC. The substantial short-term variability in emissions due to the COVID-19 lockdown creates an unmistakable signal despite known model uncertainties and satellite observation errors, providing a unique opportunity to develop and validate satellite-based carbon emissions monitoring.

The uncertainties in our results lie in satellite observations of $NO_2$ column densities, determination of the $NO_2$ column response to surface $NO_x$ emission changes by the GEOS-Chem model, differentiating sectoral $NO_x$ emissions from satellite-constrained estimates, as well as $CO_2$ to $NO_x$ emission ratios. Using spatial-temporal average and the relative difference of TROPOMI data is expected to cancel a major part of the systematic errors in satellite observations. The robustness of the response factor between $NO_2$ column and anthropogenic $NO_x$ emissions is proved by a series of sensitivity runs of the GEOS-Chem model. Corrections on sectoral emissions are also tested through different definitions of the grid-cell dominant source sector and the inversion results of ten-day mean $CO_2$ emissions are found to be stable (Fig. S12). The influence of $CO_2$ to $NO_x$ ratios on the estimates of $CO_2$ relative reductions from the 2019 MEIC data to the 2020 inversions is marginal as we use the consistent sector-specific ratio maps with MEIC. The robustness of our estimates could also be demonstrated by the broad consistency of our estimates with previous studies *(6, 13)* and with independent economic and industrial statistics data collected on the monthly scales (Fig. S14). More details about the uncertainty analysis could be found in *Supplementary Materials*.

Our findings show that although the COVID-19 lockdown dramatically reduced China's $CO_2$ emissions during the first two months after the virus outbreak, the emissions rebounded quickly as lockdown measures were lifted. In the first month after Wuhan reopened, which essentially marked the end of the country lockdown, China's $CO_2$ emissions again reached pre-COVID-19 levels and exceeded the level of emissions in the same period of 2019. These trends in emissions were strongly influenced by the industry sector, which was the main driver in both the decline and rebound of Chinese $CO_2$ emissions over the short period. It is difficult to predict the long-term effect of the pandemic on energy use and thus $CO_2$ emissions. Lockdowns in some other regions are now being relaxed, and economic stimulus packages are now being implemented in China and other countries affected by COVID-19. But our results show that the effects on energy and emissions depend on the strictness and duration of lockdowns in different regions, the economic structures and recovery stimulus programs of those regions, the effectiveness of such stimulus, and also whether and where a second wave of the virus and lockdowns recur. Regardless, the method we have developed and demonstrate here will support efforts to monitor changes in energy systems and emissions in near real-time as we navigate into that uncertain future.

**References and Notes**

1. H. Tian *et al.*, An investigation of transmission control measures during the first 50 days of the COVID-19 epidemic in China. *Science* **368**, 638-642 (2020).
2. IEA, Global Energy Review 2020. *IEA*, Paris https://www.iea.org/reports/global-energy-review-2020 (2020).
3. F. Liu *et al*., Abrupt declines in tropospheric nitrogen dioxide over China after the outbreak of COVID-19. Preprint at https://arxiv.org/abs/2004.06542v1 (2020).
4. Z. S. Venter *et al*., COVID-19 lockdowns cause global air pollution declines with implications for public health risk. Preprint at https://doi.org/10.1101/2020.04.10.20060673 (2020).
5. M. Bauwens *et al.*, Impact of coronavirus outbreak on $NO_2$ pollution assessed using TROPOMI and OMI observations. *Geophys. Res. Lett.*, e2020GL087978 (2020).

6. C. Le Quéré *et al.*, Temporary reduction in daily global $CO_2$ emissions during the COVID-19 forced confinement. *Nature Climate Change*, (2020).

7. G. Janssens-Maenhout *et al.*, EDGAR v4.3.2 Global Atlas of the three major greenhouse gas emissions for the period 1970–2012. *Earth Syst. Sci. Data* **11**, 959-1002 (2019).

8. G. P. Peters *et al.*, Rapid growth in $CO_2$ emissions after the 2008–2009 global financial crisis. *Nature Climate Change* **2**, 2-4 (2012).

9. M. McKee, D. Stuckler, If the world fails to protect the economy, COVID-19 will damage health not just now but also in the future. *Nature Medicine* **26**, 640-642 (2020).

10. D. Guan *et al.*, Global supply-chain effects of COVID-19 control measures. *Nature Human Behaviour*, (2020).

11. P. Friedlingstein *et al.*, Global Carbon Budget 2019. *Earth Syst. Sci. Data* **11**, 1783-1838 (2019).

12. J. Tollefson, How the coronavirus pandemic slashed carbon emissions — in five graphs. *Nature* (2020).

13. Z. Liu *et al.*, COVID-19 causes record decline in global $CO_2$ emissions. Preprint at https://arxiv.org/abs/2004.13614 (2020).

14. A. Eldering *et al.*, The Orbiting Carbon Observatory-2 early science investigations of regional carbon dioxide fluxes. *Science* **358**, eaam5745 (2017).

15. F. M. Schwandner *et al.*, Spaceborne detection of localized carbon dioxide sources. *Science* **358**, eaam5782 (2017).

16. S. S. Basu *et al.*, Global $CO_2$ fluxes estimated from GOSAT retrievals of total column $CO_2$. *Atmos. Chem. Phys.* **13**, 8695-8717 (2013).

17. R. Nassar *et al.*, Quantifying $CO_2$ Emissions From Individual Power Plants From Space. *Geophysical Research Letters* **44**, 10,045-010,053 (2017).

18. B. Zheng *et al.*, Observing carbon dioxide emissions over China's cities with the Orbiting Carbon Observatory-2. *Atmos. Chem. Phys. Discuss.* **2020**, 1-17 (2020).

19. J. P. Veefkind *et al.*, TROPOMI on the ESA Sentinel-5 Precursor: A GMES mission for global observations of the atmospheric composition for climate, air quality and ozone layer applications. *Remote Sensing of Environment* **120**, 70-83 (2012).

20. R. V. Martin *et al.*, Global inventory of nitrogen oxide emissions constrained by space-based observations of NO2 columns. *Journal of Geophysical Research: Atmospheres* **108**, (2003).

21. S. Beirle, K. F. Boersma, U. Platt, M. G. Lawrence, T. Wagner, Megacity Emissions and Lifetimes of Nitrogen Oxides Probed from Space. *Science* **333**, 1737-1739 (2011).

22. B. Mijling, R. J. van der A, Using daily satellite observations to estimate emissions of short-lived air pollutants on a mesoscopic scale. *Journal of Geophysical Research: Atmospheres* **117**, (2012).

23. S. Beirle *et al.*, Pinpointing nitrogen oxide emissions from space. *Science Advances* **5**, eaax9800 (2019).

24. E. V. Berezin *et al.*, Multiannual changes of $CO_2$ emissions in China: indirect estimates derived from satellite measurements of tropospheric $NO_2$ columns. *Atmos. Chem. Phys.* **13**, 9415-9438 (2013).
25. I. B. Konovalov *et al.*, Estimation of fossil-fuel $CO_2$ emissions using satellite measurements of "proxy" species. *Atmos. Chem. Phys.* **16**, 13509-13540 (2016).
26. F. Liu *et al.*, A methodology to constrain carbon dioxide emissions from coal-fired power plants using satellite observations of co-emitted nitrogen dioxide. *Atmos. Chem. Phys.* **20**, 99-116 (2020).
27. M. Reuter *et al.*, Decreasing emissions of $NO_x$ relative to $CO_2$ in East Asia inferred from satellite observations. *Nature Geoscience* **7**, 792-795 (2014).
28. B. Zheng *et al.*, Trends in China's anthropogenic emissions since 2010 as the consequence of clean air actions. *Atmos. Chem. Phys.* **18**, 14095-14111 (2018).
29. L. N. Lamsal *et al.*, Application of satellite observations for timely updates to global anthropogenic $NO_x$ emission inventories. *Geophys. Res. Lett.* **38**, (2011).
30. Y. X. Wang, M. B. McElroy, D. J. Jacob, R. M. Yantosca, A nested grid formulation for chemical transport over Asia: Applications to CO. *Journal of Geophysical Research: Atmospheres* **109**, (2004).

**Acknowledgments**

The authors acknowledge the free use of tropospheric $NO_2$ column data from the TROPOMI sensor from www.temis.nl. **Funding:** This work was supported by the National Natural Science Foundation of China (41921005 and 41625020). **Author contributions:** Conceptualization and Methodology: Q. Z., B. Z., G. G.; Formal analysis and Investigation: B. Z., G. G., Q. Z.; Writing - Original Draft: B. Z., G. G., S. J. D.; Writing - Review & Editing: all; Visualization: B. Z., G. G., Q. Z., N. W.; Supervision and Project administration: Q. Z.; Funding acquisition: Q. Z., D. G.. **Competing interests:** The authors declare no competing interests. **Data and materials availability:** Codes used in data analysis are available from the corresponding author upon reasonable request.

**Supplementary Materials**

Materials and Methods

Figures S1-S14

Table S1

References (*1-33*)

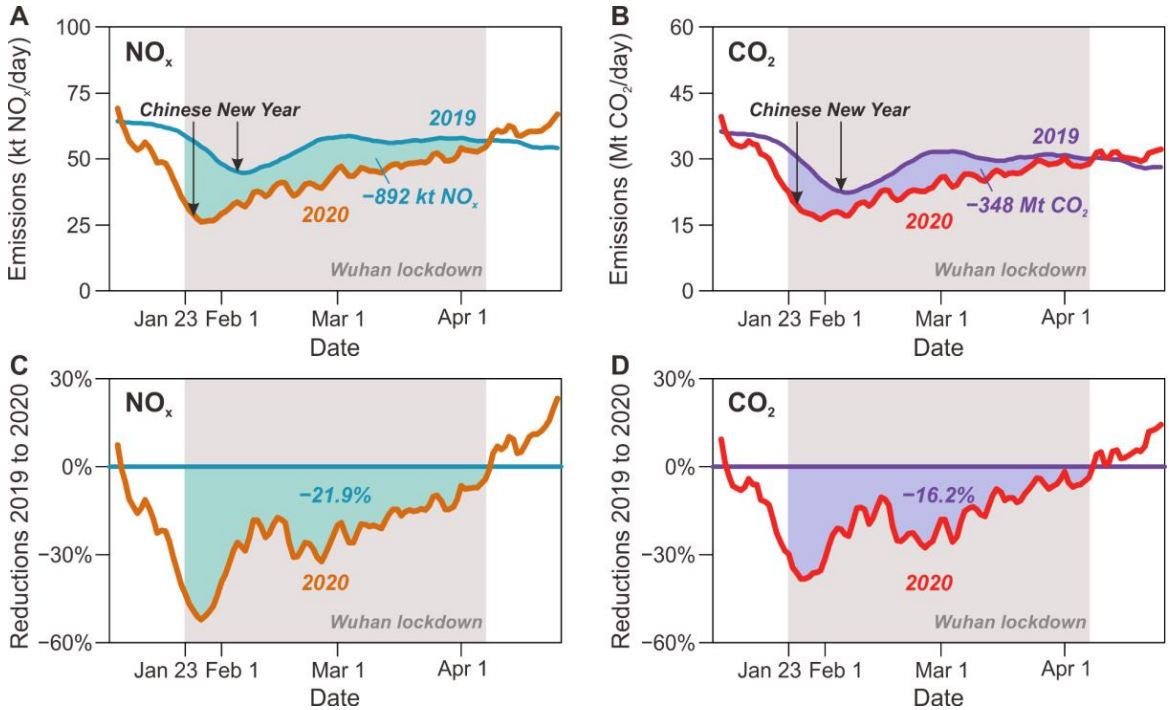

**Fig. 1. Ten-day moving average of NO$_x$ and CO$_2$ emissions from January to April in 2019 and 2020.** The 2019 emissions data are derived from the MEIC emission inventory, and the 2020 emission data are the TROPOMI-constrained emission estimate from this study. For NO$_x$ ((**A**) and (**C**)), the yellow and green curves are for 2020 and 2019, respectively, both of which are plotted according to the date in the x-axis. The y-axis in (**A**) and (**C**) represent the ten-day moving average of NO$_x$ emissions and the NO$_x$ emissions reductions from 2019 to 2020, respectively. The figures (**B**) and (**D**) are plotted for the CO$_2$ emissions, which are similar to (**A**) and (**C**), respectively. The grey shades reflect the period of the Wuhan lockdown from 23rd January 2020 to 7th April 2020.

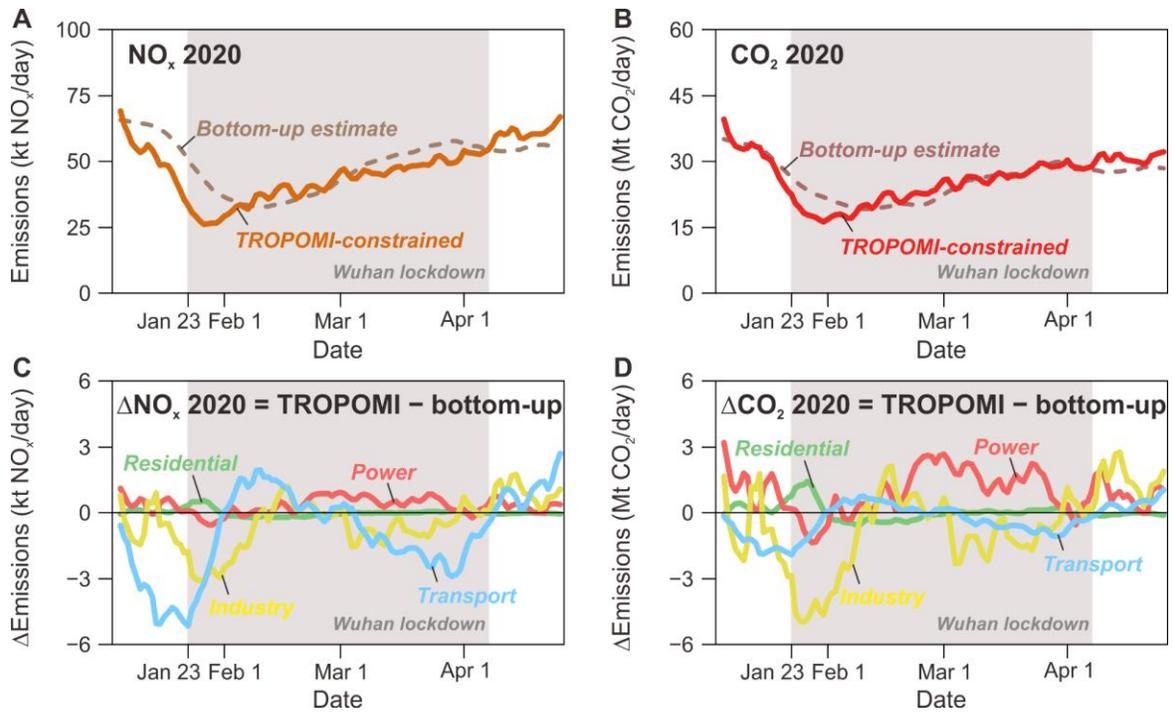

**Fig. 2. Comparisons between TROPOMI constrained and bottom-up emission estimates.** The solid curves in (**A**) and (**B**) are the ten-day moving average of TROPOMI-constrained $NO_x$ and $CO_2$ emissions, respectively, from January 2020 to April 2020. The dashed curves in (**A**) and (**B**) are the bottom-up estimates. The difference of the ten-day moving average emissions between the TROPOMI-based and bottom-up estimates are plotted for $NO_x$ in (**C**) and plotted for $CO_2$ in (**D**) for the source sectors of power (red), industry (yellow), residential (green), and transport (blue).

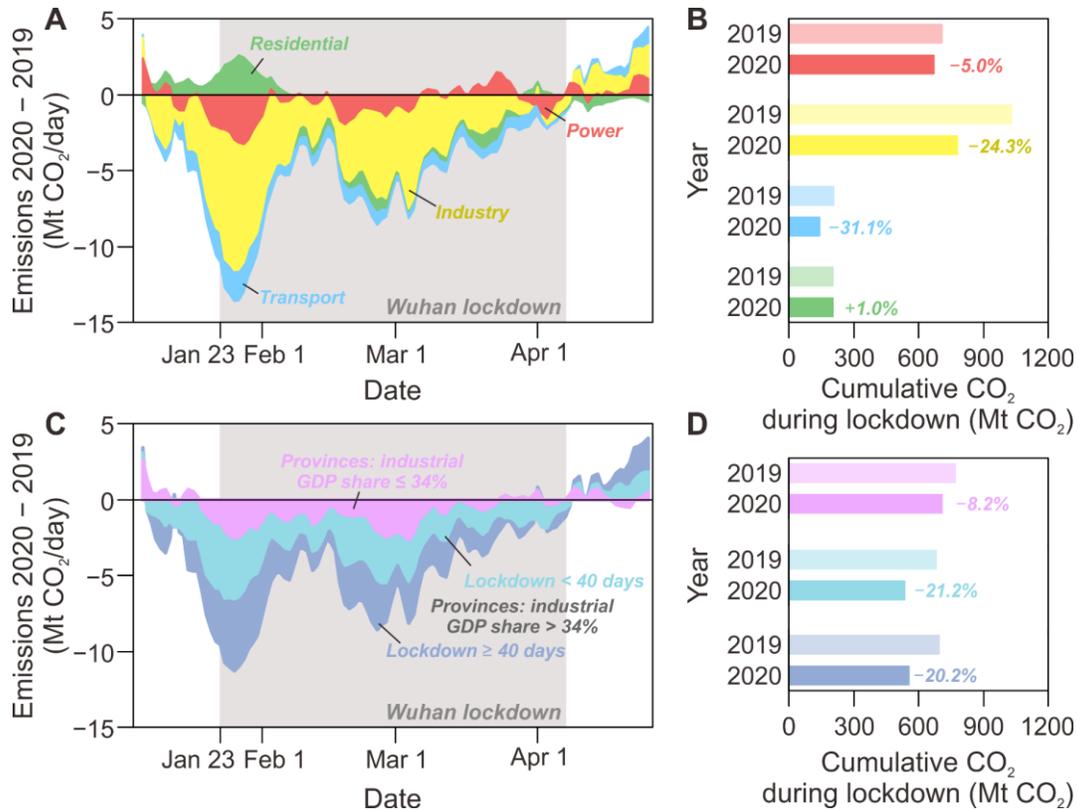

**Fig. 3. Decomposition of the difference in the ten-day moving average of $CO_2$ emissions between 2019 and 2020 by source sector and by source region.** The emissions difference is split into power, industry, residential, and transport sectors in (**A**), and split into three regional categories in (**C**), including 1) the provinces with the share of industrial GDP in provincial total GDP lower than 34%, 2) the provinces with a share of industrial GDP higher than 34% and a lockdown shorter than 40 days, and 3) the provinces with the industrial GDP share higher than 34% and a lockdown longer than 40 days. The cumulative $CO_2$ emissions during Wuhan lockdown (grey shades in (**A**) and (**C**)) are presented by source sector in (**B**) and by source region in (**D**).

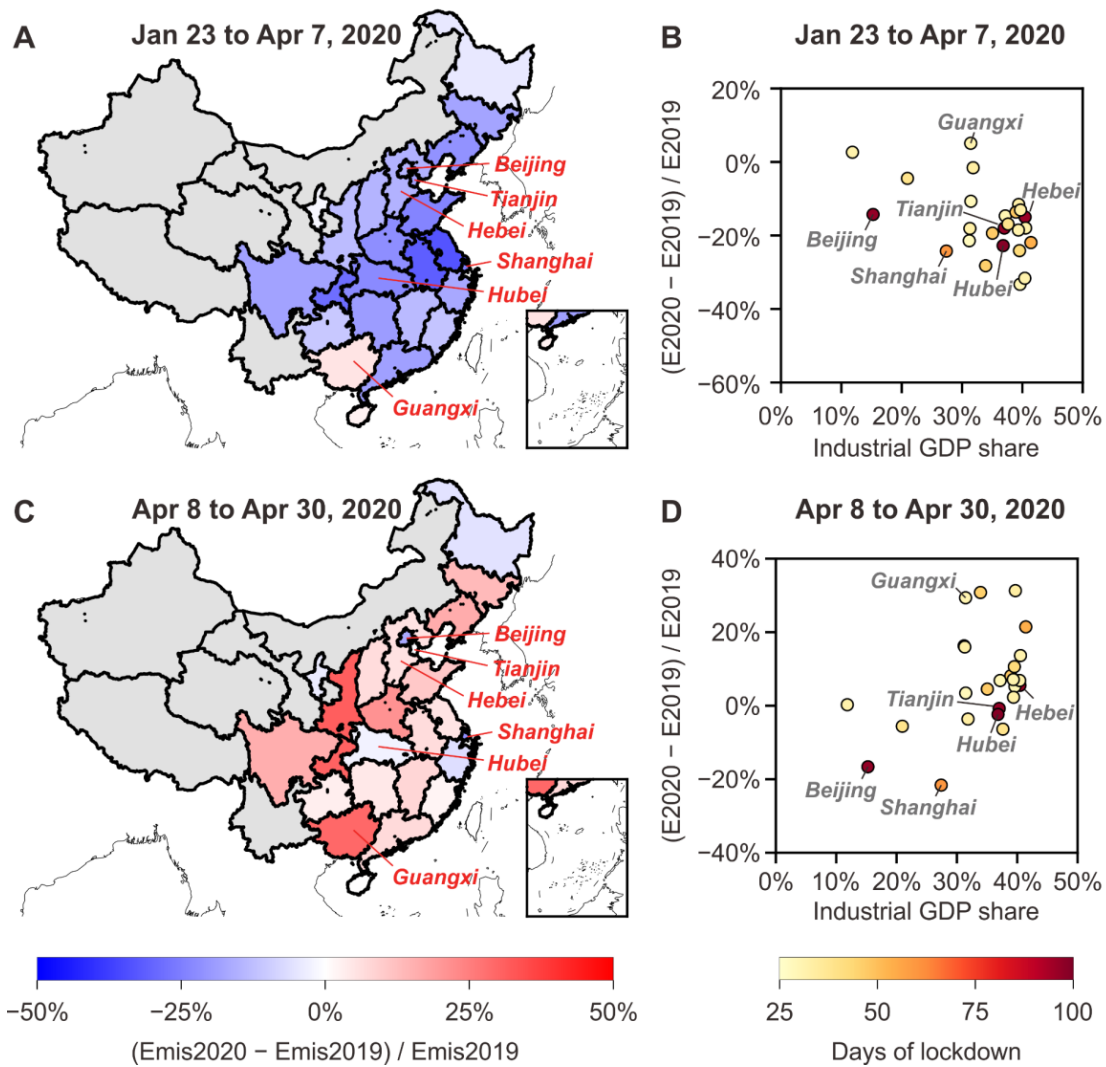

**Fig. 4. Relative changes in provincial CO$_2$ emissions of mainland China from 2019 to 2020.** The relative changes in the cumulative CO$_2$ emissions between 23rd January and 7th April are presented in (**A**) and (**B**). We only plot the provinces with enough TROPOMI observations used in this study (NO$_2$ TVCDs larger than $1\times10^{15}$ molec/cm$^2$) that covered more than 80% of their anthropogenic NO$_x$ emissions from January to April in 2019 and 2020. The provinces without enough TROPOMI observations are plotted with the grey color. The colors of other provinces in (**A**) represent the magnitude of the cumulative CO$_2$ emissions changes comparing the same period between 2019 and 2020. Each dot in (**B**) represents a province in mainland China, which is plotted based on the share of industrial GDP in the provincial total GDP along the x-axis and the changes in cumulative CO$_2$ emissions along the y-axis. The color of each dot represents the number of days when each province remained its public health response system to the COVID-19 emergency at the top level. Additional lines point to the provinces of Beijing, Tianjin, Hebei, Shanghai, Hubei, and Guangxi. Figures (**C**) and (**D**) are similar to (**A**) and (**B**), respectively, but present results for the period from 8th April to 30th April.

# Supplementary Materials for

# Satellite-based estimates of decline and rebound in China's CO₂ emissions during COVID-19 pandemic

Bo Zheng, Guannan Geng, Philippe Ciais, Steven J. Davis, Randall V. Martin, Jun Meng, Nana Wu, Frederic Chevallier, Gregoire Broquet, Folkert Boersma, Ronald van der A, Jintai Lin, Dabo Guan, Yu Lei, Kebin He, Qiang Zhang

Correspondence to: Qiang Zhang (qiangzhang@tsinghua.edu.cn)

**This PDF file includes:**

    Materials and Methods
    Figs. S1 to S14
    Table S1

## Materials and Methods

**Integrated modeling framework**

This work integrates a detailed bottom-up inventory, high-quality spaceborne constraints, and a chemical transport model to estimate 10-day mean dynamic changes in China's $NO_x$ and $CO_2$ emissions from January to April in 2020, as depicted in Figure S1. We first develop a preliminary bottom-up estimate of daily $NO_x$ and $CO_2$ emissions in 2020 based on the Multi-resolution Emission Inventory for China (MEIC, http://www.meicmodel.org/) (*1, 2*) in 2019 and relative changes in activity data and emission factors from 2019 to 2020. We then use a nested-grid GEOS-Chem model (*3, 4*) at the horizontal resolution of 0.5° × 0.625° to establish the spatial-temporal varied local relationship between the changes in surface $NO_x$ emissions and changes in $NO_2$ tropospheric vertical column densities (TVCDs). Such a relationship is applied to changes in the satellite observed $NO_2$ TVCDs from 2019 to 2020 derived from the TROPOspheric Monitoring Instrument (TROPOMI) (*5*) to infer high-resolution $NO_x$ emissions in 2020 on the ten-day moving average scale. Prior to the top-down emission estimation, changes in satellite $NO_2$ TVCDs contributed by meteorological variations have been excluded. Next, we use the top-down estimates of $NO_x$ emissions in 2020 to correct the sectoral emission dynamics in the bottom-up emission estimates in 2020, based on the emission differences revealed by the grid cells dominated by a single source sector. Correction in the emission sectoral information helps capture the dynamics of the $CO_2$ to $NO_x$ emission ratio in 2020, which is different from that in 2019. Finally, the TROPOMI-constrained $NO_x$ emissions with the constrained dynamic sectoral emission changes are combined with sector-specific ratios of $CO_2$ to $NO_x$ emissions to infer the ten-day moving average of $CO_2$ emissions in 2020 from specific sectors.

**Bottom-up estimates of daily $NO_x$ and $CO_2$ emissions**

Daily emissions in 2019

The daily emissions of China in 2019 are derived from the MEIC model developed by Tsinghua University (*1, 2*). MEIC is a technology-based emission model that tracks the evolution of the manufacturing and pollution control technologies for more than 700 anthropogenic emission sources in China, which can be integrated into five source sectors of power, industry, residential, transportation, and agriculture. The coal-fired power plants in the power sector (*6, 7*) and the cement plants in the industry sector (*8*), the two important sources of $NO_x$ and $CO_2$ emissions in China, are treated as point sources with accurate geographic locations, where emissions are estimated using the facility-level activity data and emission factors. The other emission sources are all estimated as area sources using the province- or county-level parameters (*9, 10*), and the emissions are downscaled on the grid cells at the horizontal resolutions of 2° × 2.5° and 0.5° × 0.625°, respectively, to run the GEOS-Chem global and nested grid simulations. At such spatial resolutions, the MEIC emission inventory has an accurate representation of emission spatial distributions (*11, 12*) and emission annual trends (*13, 14*), evaluated by satellite observations.

The MEIC model uses monthly and daily temporal profiles including electricity generation, industry factory operating rate, traffic volume index, and heating degree day to disaggregate the annual emission estimates to daily emissions. The daily sectoral $NO_x$ emissions from January to April in 2019 are shown in Figure S2a. The $CO_2$ to $NO_x$ emission ratio varies substantially by source sector. According to the MEIC model results in 2019, the $CO_2$ to $NO_x$ emission ratios are on average 979, 623, 917, and 141 for the source sectors of power, industry, residential, and transport, respectively. The largest value for the power sector accounts for the pollution control on $NO_x$ in the coal-fired power plants, which has reduced the $NO_x$ emission factors significantly. The large value for the residential sector is because the residential stoves with low combustion efficiency generate less $NO_x$ per fuel burnt than the high-efficiency boilers used in the power and industry sectors. The low $CO_2$ to $NO_x$ emission ratio of the transport sector is due to low carbon content of petroleum-based fuels and the high emission factors of $NO_x$ from diesel vehicles and nonroad equipment. The $CO_2$ to $NO_x$ emission ratios distinct between source sectors combined with the daily emission variations in these sectors further drive the daily dynamics of the national average $CO_2$ to $NO_x$ emission ratio as shown in Figure S2b. The low values around the Chinese New Year in 2019 are due to the reduced activities in the power and industry sectors during the holiday.

Daily emissions in 2020

The daily emissions from January to April in 2020 are estimated on the base of the 2019 MEIC emissions data combined with the relative changes in activity data and emission factors from 2019 to 2020 according to the method of (*15*). The 2020 emissions are estimated based on the four source sectors in MEIC, which are power, industry, residential, and transportation. The monthly emissions in 2020 are estimated first and are then disaggregated onto the daily time scales.

The monthly growth rates of electricity generation, cement production, iron production, and industrial Gross Domestic Product (GDP) from 2019 to 2020 are used to represent the relative changes in the activity data of power plants, cement plants, iron and steel plants, and the other industries, respectively. The monthly statistics data are derived from China's National Bureau of Statistics (http://www.stats.gov.cn/). It should be noted that several assumptions had to be made to make the emission estimates in 2020 possible. First, the statistics data for January and February were combined together, so we assume that the activity changes in these two months are the same. Second, the growth rate in the electricity generation is only available for all of the thermal electricity, which includes coal-, oil-, and natural gas-fired electricity that cannot be further split. We assume that the activity data of all of the thermal power plants follow the same monthly changes from 2019 to 2020. Third, the industrial GDP is used to roughly represent the relative change of the industrial activities, although the economic values may not directly reflect the changes in the fossil fuel burned. These three assumptions mentioned above could cause uncertainties in the estimates of monthly emissions in the power and industrial sectors in 2020.

For the residential and transport sectors, there are no monthly statistics of how much fuel has been burned in each month of 2020. We rely on the population-weighted heating degree day (HDD) and on the Baidu population migration data, respectively, to represent the relative changes in the activities of these two sectors from 2019 to 2020. For the residential sector, we calculate the population-weighted HDD for all of China's cities in each month of 2019 and 2020 to represent the relative change of the residential energy use in each city. The gridded population data are exploited from Gridded Population of the World, Version 4 (GPWv4) (*16*), and the surface temperature data are derived from the ERA5 dataset produced by ECMWF (*17*). The reference temperature is set at 18 °C (*18*). For the transport sector, we use the Baidu migration data (https://qianxi.baidu.com/2020/) within each city to represent the relative change in activity data of the transport sector from 2019 to 2020 in each city. There could be uncertainties in using Baidu migration data to estimate transport emissions because the Baidu location-based services track all of the surface transport within cities including walk, bicycles, vehicles, and subways. And the way we use this index to predict activity changes in both the road and nonroad sources also involves uncertainties.

We multiply the monthly growth rates in the activity data of each emission source sector mentioned above by the emissions in the corresponding month of 2019 and the change of $NO_x$ emission factors from the beginning to the end of 2019 given by the MEIC model to estimate the monthly $NO_x$ emissions from January to April in 2020. The emission factor of $CO_2$ is assumed unchanged so the monthly $CO_2$ emissions are estimated just based on the activity growth rates. The monthly emissions in 2020 are split into the daily time scales based on several proxies. The emissions from power plants are split into daily scales based on the daily coal consumption of the six major power generation groups in China (Fig. S3a), which accounted for 12% of China's coal consumption in the power sector in 2019. The industrial emissions from iron and steel plants, coke plants, and coal mines are split into daily scales based on the daily operating rates over the whole of China (Fig. S3b to S3e). All the other industrial emissions are split into the daily scale on the base of daily coal consumption in the major power generation groups (Fig. S3a), which causes uncertainties in the daily emission estimates of the industry sector. The monthly emissions from the residential and transport sectors of each city are split into daily emissions based on the daily variation of the population-weighted HDD (Fig. S4) and the Baidu index (Fig. S5) in each city.

**10-day mean $NO_x$ emissions inferred from TROPOMI**

Relative changes in $NO_2$ TVCDs between 2019 and 2020 from TROPOMI

Relative changes in $NO_2$ TVCDs over China are retrieved from TROPOMI (*5*). TROPOMI is the single sensor onboard the European Copernicus Sentinel-5 Precursor (S5P) satellite launched in 2017. The instrument provides nearly daily global coverage and ascends across the equator at approximately 13:30 local time. We use $NO_2$ TVCDs from the official TM5-MP-DOMINO version 1.2/1.3 offline product (http://www.temis.nl/airpollution/no2col/data/tropomi/) (*19*). Only pixels with quality assurance value above 0.5 and cloud fraction below 30% are kept to reduce the retrieval errors. $NO_2$ measurements are then aggregated to a spatial resolution of $0.5° \times 0.625°$ to match the grid of GEOS-Chem (Fig. S6a and S6b). During the first four months of 2020, the satellite $NO_2$ TVCDs decreased in most places over China, especially in Wuhan and the surrounding area (Fig. S6c). On average, the national $NO_2$ TVCDs drop over China is 22% compared to the same period in 2019.

The unprecedented spatial coverage and resolution of TROPOMI retrievals enhance the capability of tracking the changes in satellite $NO_2$ TVCDs on a high spatial-temporal scale (*20, 21*). To smooth out daily fluctuations in $NO_2$ TVCDs caused by random errors and increase the sample number and spatial coverage, we calculate the 10-day moving average in each year (Fig. S6e). Figure S7 shows the distribution of the valid sample days in each grid cell for each 10-day period. Note that grid cells dominated by natural sources (defined as $NO_2$ TVCDs below $1 \times 10^{15}$ molec/cm$^2$ in this study), which are more than half of all grid cells in China, are excluded in the following analysis because we mainly focus on anthropogenic emissions (*22*). In the remaining valid grid cells, the average share of grid cells with more than 5 valid sample days could reach 87%, and they cover 81% of national anthropogenic $NO_x$ emissions over China (Fig. S7c), indicating that the 10-day moving average value used in this work are representative for the emission estimates over China.

Dynamic local relationship between changes in $NO_x$ emissions and $NO_2$ TVCDs provided by GEOS-Chem

$NO_2$ TVCDs are strongly related to surface $NO_x$ emissions since 1) the lifetime of $NO_x$ is short and 2) $NO_2/NO_x$ ratio is high in the boundary layer. Previous studies have used chemical transport models to build the relationship between changes in surface $NO_x$ emissions (*E*) and changes in $NO_2$ TVCDs (*Ω*) and rapidly update $NO_x$ emissions based on satellite observations (*23-26*). Following such a method, we use the nested-grid GEOS-Chem model to establish the relationship between changes in $NO_x$ emissions and $NO_2$ TVCDs.

We use the GEOS-Chem version 12.3.0 (https://doi.org/10.5281/zenodo.2620535) driven by assimilated meteorological fields from the NASA Global Modeling and Assimilation Office's Modern-Era Retrospective analysis for Research and Applications Version 2 (MERRA-2) (*27*). For the simulation over China, we use the nested-grid configuration over Southeast Asia at a spatial resolution of 0.5° × 0.625° (*3, 4*), with boundary conditions adopted from a 2° × 2.5° global simulation. We use the "tropchem" mechanism that simulates full chemistry in the troposphere. Vertical mixing in the planetary boundary layer is simulated using a nonlocal mixing scheme (*28*).

For the baseline simulation in 2019, anthropogenic emissions over Southeast Asia are taken from the MIX inventory (*29*), while emissions in mainland China are replaced by emissions from the MEIC model for the year 2019 as described in the previous section. GEOS-Chem includes additional $NO_x$ emission sources such as lightning (*30*), soil and fertilizer (*31*), shipping and aircraft. Figure S8 shows the comparison between coincidently sampled simulated and satellite-retrieved $NO_2$ TVCDs for the baseline scenario. The spatial pattern of the baseline simulation is very similar to the $NO_2$ TVCDs from TROPOMI, and they agree very well with $R^2 = 0.89$ and slope = 0.98.

Following Lamsal et al. (*23*), we also conduct a perturbation scenario with anthropogenic $NO_x$ emissions over China reduced by 40%. Then we get:

$$\beta_{t,i} = \frac{\Delta E}{E} \div \frac{\Omega_{t,i,pertubed,2019} - \Omega_{t,i,base,2019}}{\Omega_{t,i,base,2019}} \tag{S1}$$

where subscript *perturbed* and *base* represent the perturbation scenario and the baseline, respectively and *t* represents each 10-day period. $\Omega_{t,i,pertubed,2019}$ and $\Omega_{t,i,base,2019}$ are model simulated $NO_2$ TVCDs on 10-day period *t* over grid cell *i* in 2019. ΔE is the 40% emission perturbation. $\beta_{t,i}$ is a unitless factor that represents the local sensitivity of changes in $NO_2$ columns to changes in $NO_x$ on 10-day period *t* in grid cell *i*.

Figure S9a shows the spatial variation of four-month averaged *β* coincidently sampled with the TROPOMI data. Generally, *β* tends to be less than one in polluted regions such as the North China Plain, the Yangtze River Delta, and the Pearl River Delta, because an increase in $NO_x$ emissions consumes OH and increases the $NO_x$ lifetime. While in clean areas where an increase in $NO_x$ emissions decreases the $NO_x$ lifetime, *β* tends to be greater than one. Figure S9b shows the 10-day moving average of national *β* over China. *β* is smaller in winter when the concentrations of OH and RO$_2$ radicals are lower, and increases in spring, which reflects longer NO$_x$ lifetime in winter time.

Sensitivity tests of $\beta$ value

Error in $\beta$ is one of the sources of uncertainty in our overall approach. Therefore, sensitivity tests are conducted to test the robustness of $\beta$ obtained from the model simulation.

Lamsal et al. (*23*) used the global GEOS-Chem model to perform several tests and proved the stability of $\beta$. For example, they find that a perturbation of 30% NO$_x$ emissions instead of 15% changes global averaged $\beta$ by <2%. Increasing anthropogenic VOCs and CO by 15% increases the global value of $\beta$ by 2.8% and 1.0%, respectively. When NO$_x$ emissions are only perturbed for a single grid cell in Ohio, $\beta$ in neighboring grids is affected by 2–6%, which indicates that $\beta$ is barely influenced by the transport of NO$_2$ from other grid cells at the resolution of 2° × 2.5°.

In this study, we use the nested-grid model with higher resolution than the global model used in Lamsal et al. (*23*), which are believed to better resolve the nonlinear NO$_x$ chemistry and heterogeneous emission sources. However, $\beta$ obtained from smaller grid cells might be more affected by the transport of NO$_2$ from neighboring grids or background area, especially during winter time when the NO$_2$ lifetime is longer. We therefore conduct a sensitivity test using the global GEOS-Chem model to simulate $\beta$ as a comparison. Results show that the national average value of $\beta$ is similar in different model resolutions, and the estimated NO$_x$ emissions are within 2.6% difference.

We also find that perturbing NO$_x$ emissions by 30% or 50% instead of 40% only changes $\beta$ by –0.7% and 0.8%. We also conduct sensitivity scenarios with an additional 20% decrease in CO and a 40% decrease in VOCs, which increase $\beta$ by 2.5% nationally. The results of the sensitivity tests show that the $\beta$ value is quite robust.

Excluding meteorological influences in NO$_2$ TVCDs changes

Variations in meteorological conditions contribute to the changes in satellite observed NO$_2$ TVCDs between 2019 and 2020, especially on short-term time scales. To exclude meteorological impacts, we conduct a model simulation with anthropogenic emissions fixed in 2019 driven by meteorological fields in 2020 (*fixemis*). Then we get:

$$\left(\frac{\Delta\Omega}{\Omega}\right)_{t,i,anth} = \frac{\Omega_{t,i,sate,2020} - \Omega_{t,i,sate,2019} \times \Omega_{t,i,fixemis,2020} \div \Omega_{t,i,base,2019}}{\Omega_{t,i,sate,2019}} \tag{S2}$$

where $\left(\frac{\Delta\Omega}{\Omega}\right)_{t,i,anth}$ is the relative changes in NO$_2$ TVCDs between 2019 and 2020 contributed by anthropogenic emission changes on 10-day period *t* in grid cell *i*. $\Omega_{t,i,sate,2019}$ and $\Omega_{t,i,sate,2020}$ are satellite NO$_2$ TVCDs retrieved from TROPOMI on 10-day period *t* in grid cell $i$. $\Omega_{t,i,fixemis,2020}$ and $\Omega_{t,i,base,2019}$ are modeled NO$_2$ TVCDs on 10-day period $t$ in grid cell $i$ for the fixed emission scenario and baseline, respectively. The relative changes between 2019 and 2020 excluding the influence from meteorological variation are shown in Figure S6d. Compared to the sharp drop of NO$_2$ TCVDs during this time, the influence of meteorological changes is minor.

10-day mean NO$_x$ emissions inferred from TROPOMI

We use the 10-day running average $\beta$ value from equation S1, changes in NO$_2$ TVCDs between 2019 and 2020 contributed by anthropogenic emissions from equation S2, and bottom-up emissions in 2019 to infer NO$_x$ emissions in 2020, as shown below:

$$E_{t,i,constrained,2020} = \left(1 + \beta_{t,i}\left(\frac{\Delta\Omega}{\Omega}\right)_{t,i,anth}\right) \times E_{t,i,bottom-up,2019} \tag{S3}$$

where $E_{t,i,constrained,2020}$ represents the TROPOMI-constrained NO$_x$ emissions in 2020 and $E_{t,i,bottom-up,2019}$ indicates the bottom-up estimates of 2019 emissions from MEIC. Here we emphasize again that we limit our calculations only to grid cells dominated by anthropogenic emission sources, which is defined as tropospheric NO$_2$ columns above $1\times10^{15}$ molec/cm$^2$.

**10-day mean CO$_2$ emissions inferred from NO$_x$ emissions**

Sectoral NO$_x$ emissions inferred from TROPOMI constraints

Since the CO$_2$ to NO$_x$ emission ratio depends on the source sector, it is important to know sectoral NO$_x$ emissions to convert TROPOMI-constrained NO$_x$ emissions to CO$_2$ emissions. Our bottom-up emission estimates in 2020 provide sectoral emissions at the daily scales, while the imperfect daily statistics data we used leave uncertainties in the sectoral emissions, which show discrepancies in total NO$_x$ emissions compared to the TROPOMI-constrained estimates (Fig. S10a). Here, we use the TROPOMI-constrained NO$_x$ emissions to correct the biases in the sectoral emission distributions from the bottom-up inventory in 2020, which finally makes the bottom-up emissions data match top-down estimates (Fig. S10a). Our method is described as follows.

First, we subtract the bottom-up estimated NO$_x$ emission map in 2020 from the TROPOMI-constrained NO$_x$ emission map at the time scale of the 10-day moving average.

$$\Delta NOx_{t,i} = NOx_{t,i,TROPOMI} - NOx_{t,i,bottom-up} \tag{S4}$$

where $t$ is the time on the base of the 10-day moving average from January to April in 2020, $i$ is the 0.5° × 0.625° grid cell, $NOx_{t,i,TROPOMI}$ is the TROPOMI-constrained NO$_x$ emissions at grid $i$ in time $t$, $NOx_{t,i,bottom-up}$ is the bottom-up estimate of NO$_x$ emissions at grid $i$ in time $t$, $\Delta NOx_{t,i}$ is the gridded emission difference between TROPOMI-constrained and bottom-up emission estimates.

Next, we classify grid cells based on the dominant emission source sector in each grid cell. Before the COVID-19 lockdown (23rd January 2020), we assume that the dominant sector in each grid cell in 2020 is consistent with that in the same period of 2019. We calculate the average percentages of emissions from power, industry, residential, and transport sectors in each grid cell using the MEIC emissions in January 2019. The dominant emission source sector in each grid cell is defined as the source sector that accounts for more than 50% of emissions in that grid. Figure S11 shows the examples of the grid cells dominated by power, industry, and transport sectors, where the TROPOMI can observe large $NO_2$ column enhancement at the locations of point sources and road networks. We also try the threshold values of 60%, 70%, and 80% to test the sensitivity, which will be discussed later. Then we calculate the differences in national emissions estimated for each source sector in 2020 (Fig. S10b).

$$\Delta NOx_{t,s} = \sum_i \Delta NOx_{t,i,s} \tag{S5}$$

where $t$ is the time before 23rd January 2020, $s$ is the source sector (power, industry, residential, and transport), $\Delta NOx_{t,i,s}$ is the emission difference between TROPOMI-constrained and bottom-up emission estimates in time $t$ at the grid cell $i$ where the $NO_x$ emissions from the sector $s$ accounts for more than 50% of emissions, $\Delta NOx_{t,s}$ is the emission difference between TROPOMI-constrained and bottom-up emission estimates in the sector $s$ for the whole country in time t. Very few grid cells are dominated by residential $NO_x$ emissions, so we do not consider this sector here.

Then, we want to know how much emissions from each emission source sector should be corrected on the basis of $\Delta NOx_{t,s}$. We calculate the contribution of the grid cells dominated by one single sector to China's total emissions of this sector. The grid cells dominated by power, industry, and transport emissions account for about 25%, 32%, and 34%, respectively, of the total $NO_x$ emissions from each sector (Fig. S10c). We rely on these values to estimate correction factors for the $NO_x$ emissions from each source sector.

$$scalefactor_{t,s} = 1 + \frac{\Delta NOx_{t,s}}{f_{t,s} \times NOx_{t,s,bottom-up}} \tag{S6}$$

$$NOx_{t,i,s,bottom-up}^{corrected} = NOx_{t,i,s,bottom-up} \times scalefactor_{t,s} \tag{S7}$$

where $t$ is the time before 23rd January 2020, $s$ the source sector (power, industry, and transport), $f_{t,s}$ is the percentage of emissions from the grid cells dominated by sector $s$ to the total emissions of this sector in time $t$, $NOx_{t,s,bottom-up}$ is the bottom-up estimate of $NO_x$ emissions from sector $s$ in time $t$, $scalefactor_{t,s}$ is the scaling factor used to correct the bottom-up estimated

NO$_x$ emissions from sector $s$ in time $t$ to get the corrected NO$_x$ emissions map for this sector ($NOx_{t,i,bottom-up}^{corrected}$).

The corrections described above are made day by day for the period before 23rd January 2020. After 23rd January 2020, the lockdown has influenced the emissions from each source sector differently, therefore the method to determine the dominant emission source sector in each grid cell cannot be performed based on the 2019 emissions anymore. Therefore for each day, we use the corrected sectoral emission maps on its prior day before we identify the dominant source sector on that day. And using the same method described above to constrain sectoral emission maps for each day.

Figure S10b shows the differences in NO$_x$ emissions between the preliminary bottom-up estimates and the TROPOMI-constrained inversions over the grid cells dominated by different source sectors. These results suggest that the uncertainties in the bottom-up estimates mainly occur in the transport and industry sectors. The preliminary bottom-up inventories in 2020 tend to overestimate emissions during the lockdown, especially at the beginning of the lockdown period, while slightly underestimate emissions after the end of lockdown. This is due to the imperfect knowledge of the changes in the daily dynamics of the transport and the industrial activities, and the proxies such as the people migration index and the industrial GDP are not accurate enough to predict emissions. The power sector shows smaller uncertainties because we have monthly electricity generation and the daily coal use in the power plants that account for 12% of China's power sector coal consumption. However, due to the uncertainties in these data as discussed before, the power sector emissions are still slightly underestimated in the bottom-up inventory.

The emissions differences shown in Figure S10b are estimated on the grid cells dominated by a single source sector, and these grid cells contribute about 20-50% of the emissions of this sector (Fig. S10c). Using this information, we have corrected the NO$_x$ emissions from power, industry, and transport sectors, while the residential emissions are not changed. We already get the daily sectoral NO$_x$ emission maps constrained by both the bottom-up information and the top-down constraints. The corrected bottom-up emission results match better with the TROPOMI-constrained emissions than the preliminary bottom-up estimates (Fig. S10a), which suggests that the grid cells dominated by different source sectors we selected represent the emission dynamics of each sector well. It should be noted that the remaining differences between the corrected bottom-up estimates and the TROPOMI inversions include the biases in the residential emissions that we ignore in the sectoral-level corrections above. Therefore, we scale the corrected sectoral bottom-up emission maps to be consistent with the TROPOMI-constrained NO$_x$ emission maps, which could partly correct the biases in the residential sector emissions and finally achieve sectoral $NO_x$ emissions maps that are consistent with the TROPOMI constrained total $NO_x$ emissions.

Sectoral $CO_2$ emissions inferred from TROPOMI constraints

We estimate the $CO_2$ emissions from the TROPOMI constrained $NO_x$ emissions according to the following equation:

$$\begin{aligned} CO2_{t,i,TROPOMI} &= \sum_s NOx_{t,i,s,TROPOMI} \times \frac{CO2_{i,s}}{NOx_{i,s}} \\ &= \sum_s NOx_{t,i,s,TROPOMI} \times \frac{A_{i,s} \times EFCO2_{i,s}}{A_{i,s} \times EFNOx_{i,s}} \\ &= \sum_s NOx_{t,i,s,TROPOMI} \times \frac{EFCO2_{i,s}}{EFNOx_{i,s}} \\ &= \sum_s NOx_{t,i,s,TROPOMI} \times \frac{EFCO2_{i,s,MEIC2019} \times (1-rCO2_{i,s})}{EFNOx_{i,s,MEIC2019} \times (1-rNOx_{i,s})} \end{aligned} \quad (S8)$$

where $CO2_{t,i,TROPOMI}$ represents the TROPOMI constrained $CO_2$ emissions at grid $i$ in time $t$, $NOx_{t,i,s,TROPOMI}$ represents the TROPOMI constrained $NO_x$ emissions from sector $s$ (power, industry, residential, and transport) at grid $i$ in time $t$, $CO2_{i,s}/NOx_{i,s}$ is the ratio of $CO_2$ to $NO_x$ emissions for sector $s$ in grid $i$, $A_{i,s}$ represents the activities of sector $s$ in grid $i$, $EFCO2_{i,s}$ and $EFNOx_{i,s}$ represent the emission factors of $CO_2$ and $NO_x$, respectively, of sector $s$ in grid $i$, $EFCO2_{i,s,MEIC2019}$ and $EFNOx_{i,s,MEIC2019}$ represent the emission factors of $CO_2$ and $NO_x$ of sector $s$ in grid $i$ derived from the MEIC 2019 data, and $rCO2_{i,s}$ and $rNOx_{i,s}$ represent the reduction ratios of $CO_2$ and $NO_x$ emission factors, respectively, from the beginning to the end of 2019.

First, we get the $CO_2$ to $NO_x$ emission ratio maps ($EFCO2_{i,s,MEIC2019}$ and $EFNOx_{i,s,MEIC2019}$) from the MEIC emission model for power, industry, residential, and transport sectors in China averaged from January to April in 2019. We adjust these emission ratio maps according to the pollution control progresses on $NO_x$ ($rNOx_{i,s}$) that occurred in China in 2019, which reduced $NO_x$ emission factors of the power and cement plants by 6–10% from 2019 to 2020. The information on the emission control measures and the impacts on the $NO_x$ emission factors are both derived from the MEIC emission model. Since the $CO_2$ emission factor of each sector is assumed stable ($rCO2_{i,s}$ equals to zero), those pollution control measures slightly increased the $CO_2$ to $NO_x$ emission ratios, which were taken into account in our analysis. The adjusted $CO_2$ to $NO_x$ emission ratio maps are multiplied by the TROPOMI-constrained sectoral $NO_x$ emissions to estimate the $CO_2$ emissions in 2020. These calculations are based on the 0.5° × 0.625° GEOS-Chen model grid cells and on the time scale of the 10-day moving average.

Figure S12 presents the $CO_2$ emissions estimates using different threshold values for the definition of the dominant source sector. We use the threshold values from 0.5 to 0.8, while the cumulative $CO_2$ emissions from January to April only vary by less than 4%. The results suggest that the identification of the dominant source sector according to different threshold values does not significantly influence the $CO_2$ emissions estimates.

We compare the ten-day moving average $CO_2$ to $NO_x$ emission ratios between 2019 and 2020 for the whole country (Fig. S13). The largest difference is that the $CO_2$ to $NO_x$ emission ratio slightly decreased during the Spring Festival holiday in 2019 while increased over the same period in 2020. This is because the transport emissions decreased substantially during the Spring Festival holiday in 2020 due to the stringent lockdown measures. The $CO_2$ to $NO_x$ emission ratios are much lower in the transport sector than the other sectors, so the average $CO_2$ to $NO_x$ emission ratios are estimated to have increased during the holiday and the lockdown period in 2020. After the lockdown measures eased, the average $CO_2$ to $NO_x$ emission ratios in 2020 returned to the same levels as in 2019. In normal years like 2019, the travel demand is still high during the Spring Festival holiday, therefore the transport emissions change much less and the $CO_2$ to $NO_x$ emission ratios decrease driven by the reduced power and industrial emissions.

**Uncertainties**

Our results are subject to a number of uncertainties and limitations. The uncertainties in estimating the top-down $NO_x$ emissions mainly come from errors in the satellite $NO_2$ TVCDs and uncertainties in $β$. The dynamic sectoral distribution of TROPOMI-constrained $NO_x$ emissions and the $CO_2$ to $NO_x$ emission ratios used to convert $NO_x$ emissions to $CO_2$ emissions introduce additional uncertainties in the final results. The uncertainty in each step of our analysis is discussed below.

First, the satellite retrievals of $NO_2$ TVCDs usually suffer from uncertainties in the radiative transfer model and in the ancillary data utilized for calculating the stratospheric $NO_2$ background and the air mass factors. For example, the TROPOMI single-pixel errors are typically ~40-60% in winter time (*32*). We use spatial and temporal averaging in our analysis to reduce the random errors. Meanwhile, TROPOMI $NO_2$ is found to be systematically underestimated over China (*33*), especially over polluted regions. However, relative differences between 2019 and 2020 are used to derive $NO_x$ emission changes, which could cancel out a major part of the systematic errors.

Second, the $β$ value simulated by GEOS-Chem model reflects the feedback of $NO_x$ emissions on $NO_x$ chemistry, which could be affected by a lot of factors, such as the model representation of chemical and physical processes (e.g. transport and deposition), the model resolution, and changes in emissions of other species involved in the $NO_x$ chemistry. We evaluate the model simulation against satellite data to prove the model's ability to capture the characteristics of $NO_2$ columns. We also conduct several sensitivity scenarios to show the robustness of $β$ value, as discussed in previous sections. The variations of $β$ when considering different perturbation magnitude of $NO_x$ emissions, influences from other species and different model resolutions are small and $β$ is proved to be quite robust.

Third, since the $CO_2$-to-$NO_x$ ratio is sector-specific, we need to capture the dynamic sectoral distribution of $NO_x$ emissions from Jan to Apr in 2020 for the conversion from $CO_2$ to $NO_x$. We use the TROPOMI-constrained $NO_x$ emissions to correct the sectoral emissions from the bottom-up estimates, based on the grid cells dominated by one single emission source, as discussed above. The thresholds used for the determination of dominant source sector in each grid could be the largest error source here. Sensitivity tests using threshold values from 0.5 to 0.8 obtain similar $CO_2$ emission estimations (Fig. S12), reflecting the robustness of our method.

Finally, the sector-specific $CO_2$-to-$NO_x$ ratio maps we achieved from the MEIC emission model also include some uncertainties, which could influence the absolute magnitude of the $CO_2$ emissions estimates. The uncertainties in $NO_x$ emission inventories are typically higher than those in $CO_2$ emissions inventories, while the MEIC $NO_x$ inventory we used in this study has revealed a good performance when modeling with the GEOS-Chem model and comparing it to the TROPOMI $NO_2$ column observations (Fig. S8). Besides, the influence of the potential uncertainties in the $CO_2$-to-$NO_x$ ratios on the estimates of $CO_2$ emissions reductions from 2019 to 2020 is relatively marginal, because we use consistent sector-specific $CO_2$-to-$NO_x$ ratio maps with the MEIC model and the comparison is also performed with the MEIC emissions.

We also compare the $CO_2$ emissions estimated from the TROPOMI-constrained $NO_x$ emissions with the economic and industrial statistics data collected on the monthly scales (Fig. S14). The relative changes in the 2020 $CO_2$ emissions compared to those in 2019 are consistent with those indicators of the industry in China, which independently evaluate our estimates of the $CO_2$ emission changes.

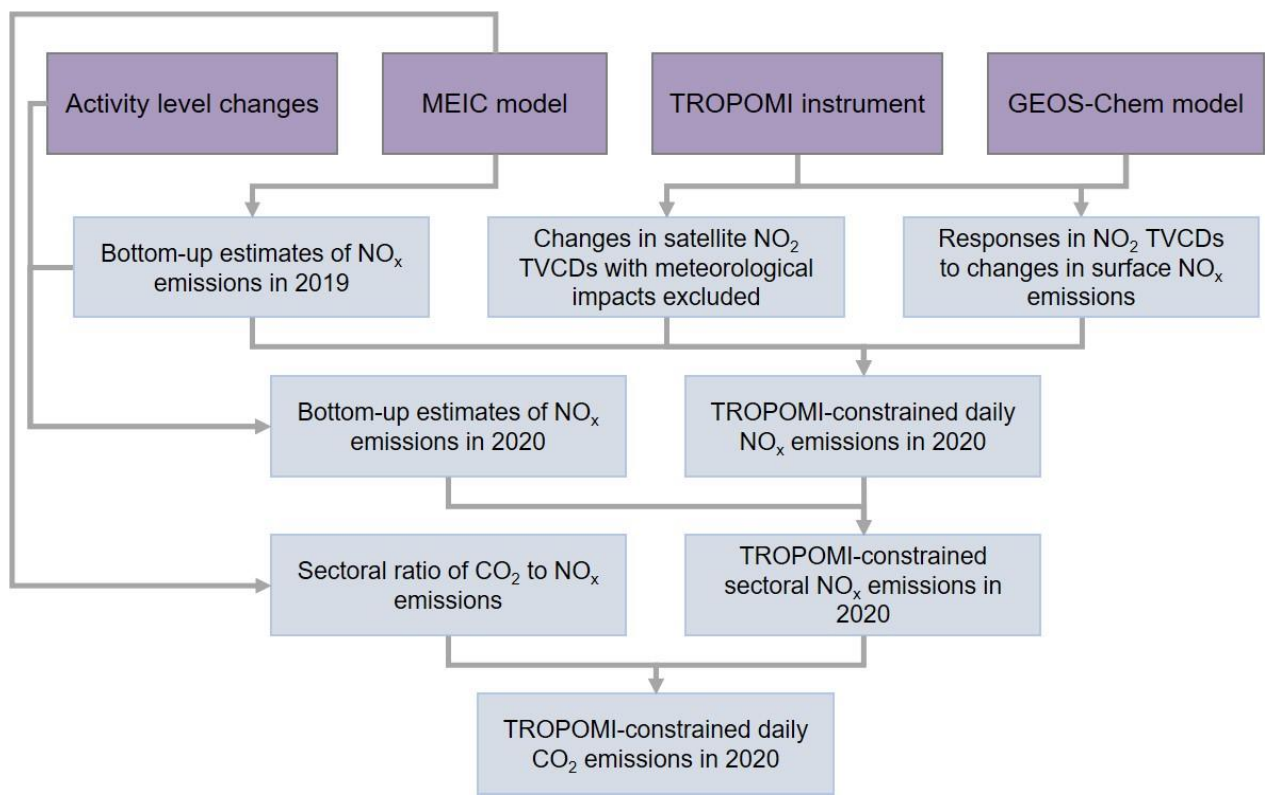

**Fig. S1. Methodology framework to infer daily CO$_2$ emissions from TROPOMI NO$_2$ TVCDs over China.**

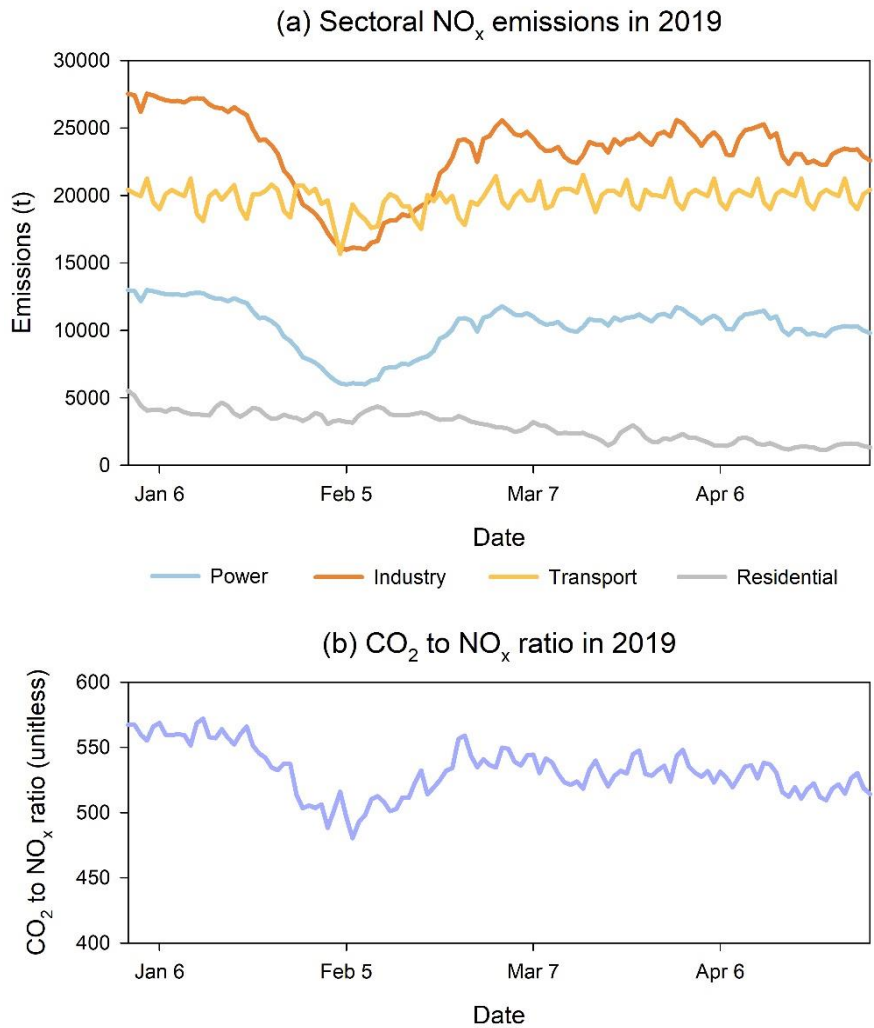

**Fig. S2. MEIC daily emissions from January to April in 2019.** Fig (**a**) presents the daily sectoral $NO_x$ emissions from the source sectors of power, industry, transport, and residential. Fig (**b**) presents the daily $CO_2$ to $NO_x$ emission ratio in 2019.

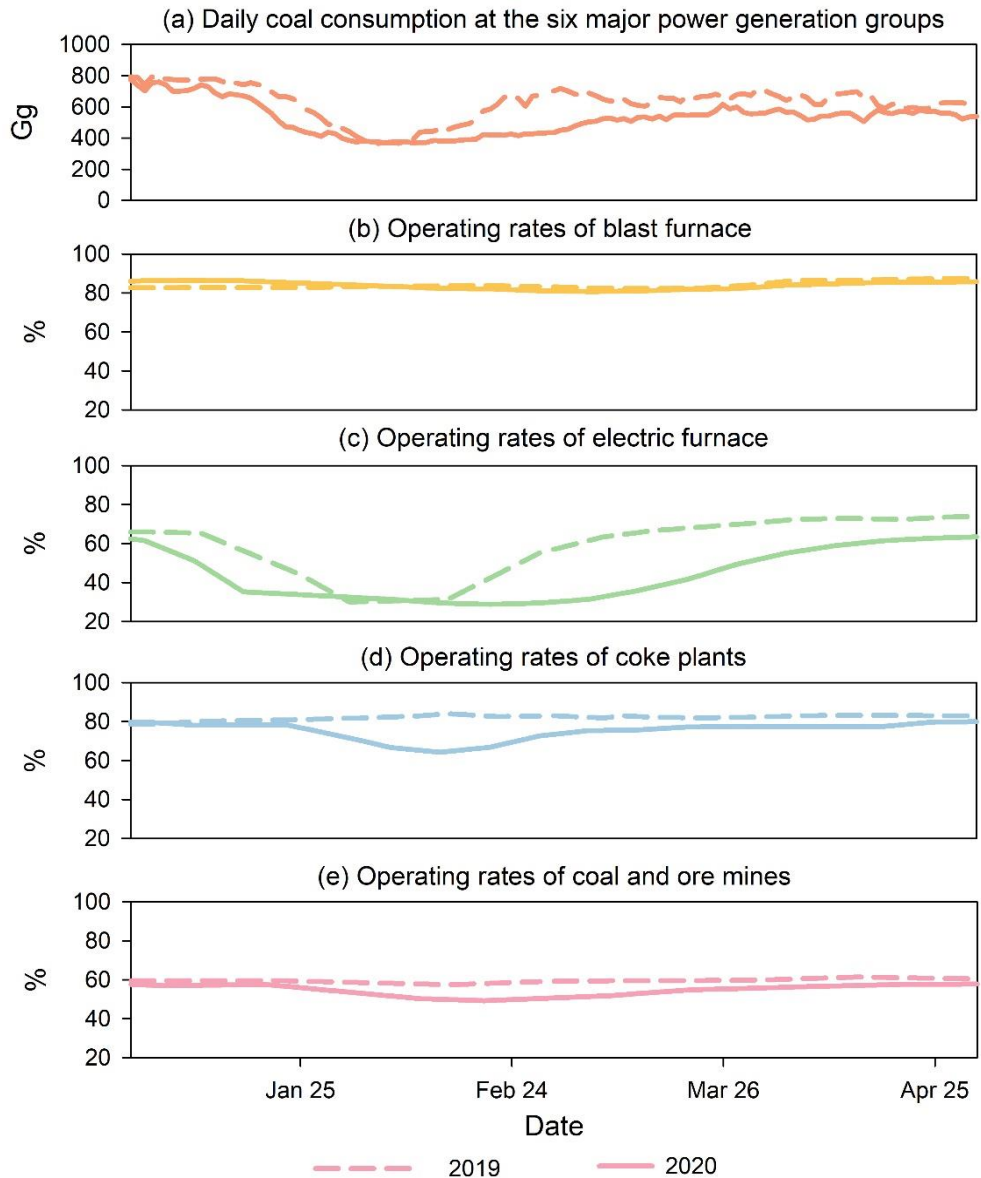

**Fig. S3. Daily activities of power and industry sectors from January to April in 2019 and 2020.** Fig. (**a**) presents daily coal consumption at the six major power generation groups. Fig. (**b**)-(**e**) present the operating rates of blast furnace, electric furnace, coke plants, and coal and ore mines.

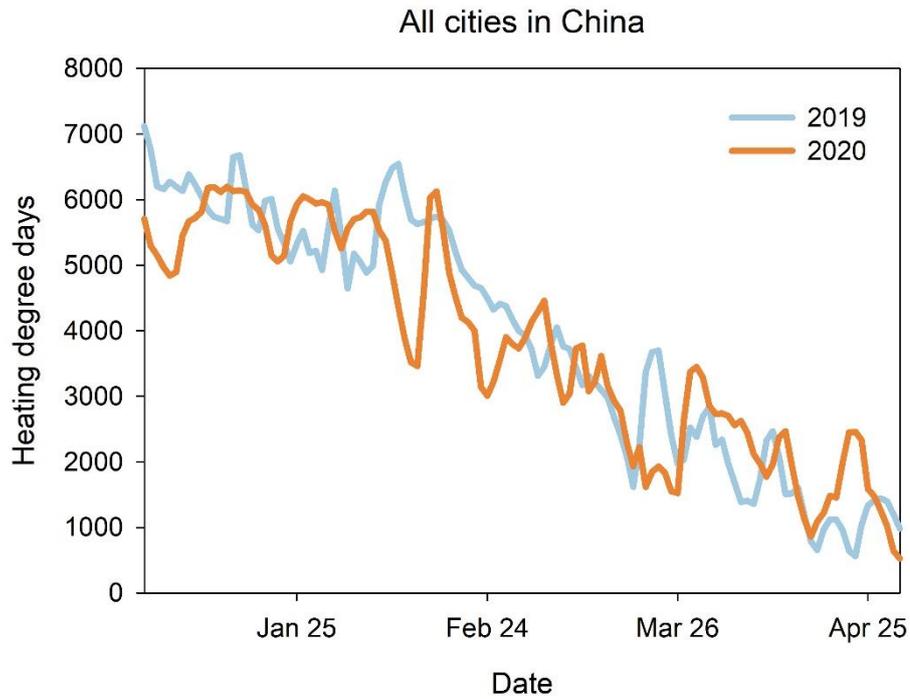

**Fig. S4. Population-weighted heating degree days of all cities in China.** The data are estimated for each city at the daily scale with the reference temperature of 18°C. The curves represent the sum of the population-weighted heating degree days of all the cities in 2019 and 2020.

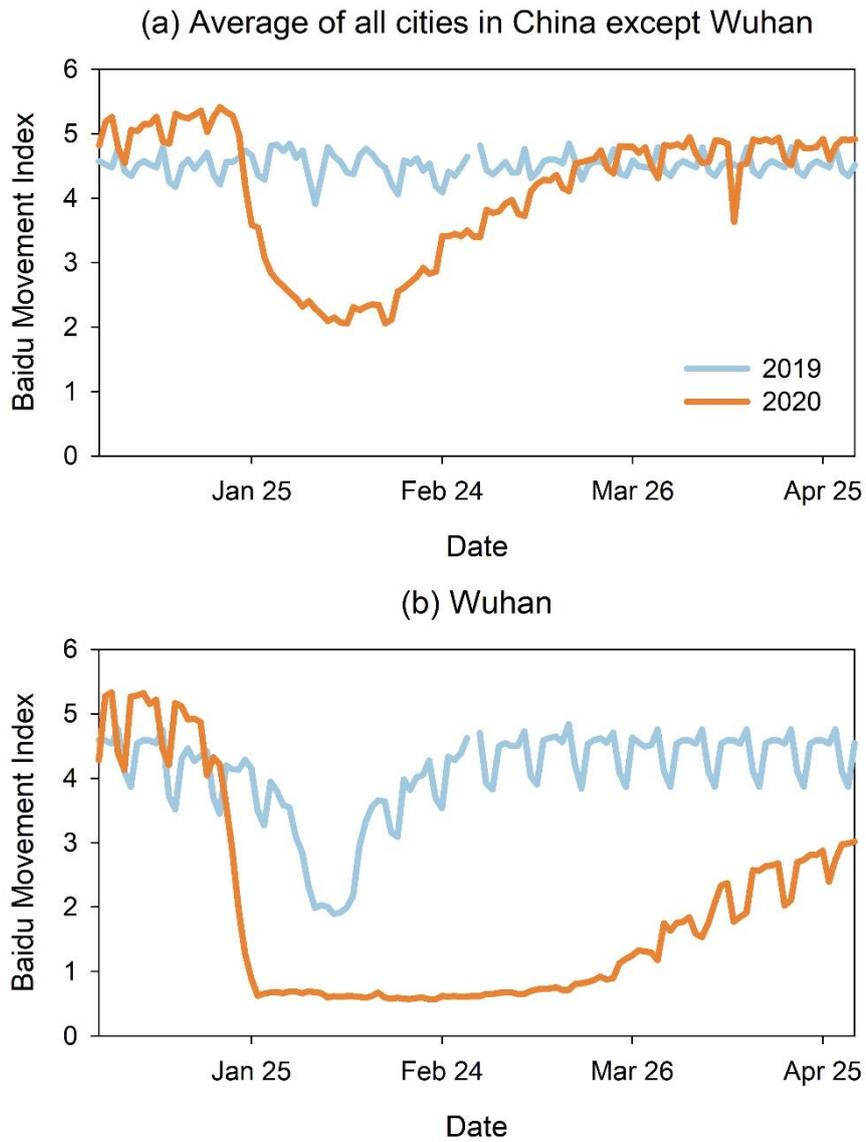

**Fig. S5. Daily population migration data within city from the Baidu location-based services.** (a) the average of all of China's cities except Wuhan, and (b) Wuhan. The Baidu migration data are derived from the website https://qianxi.baidu.com/2020/.

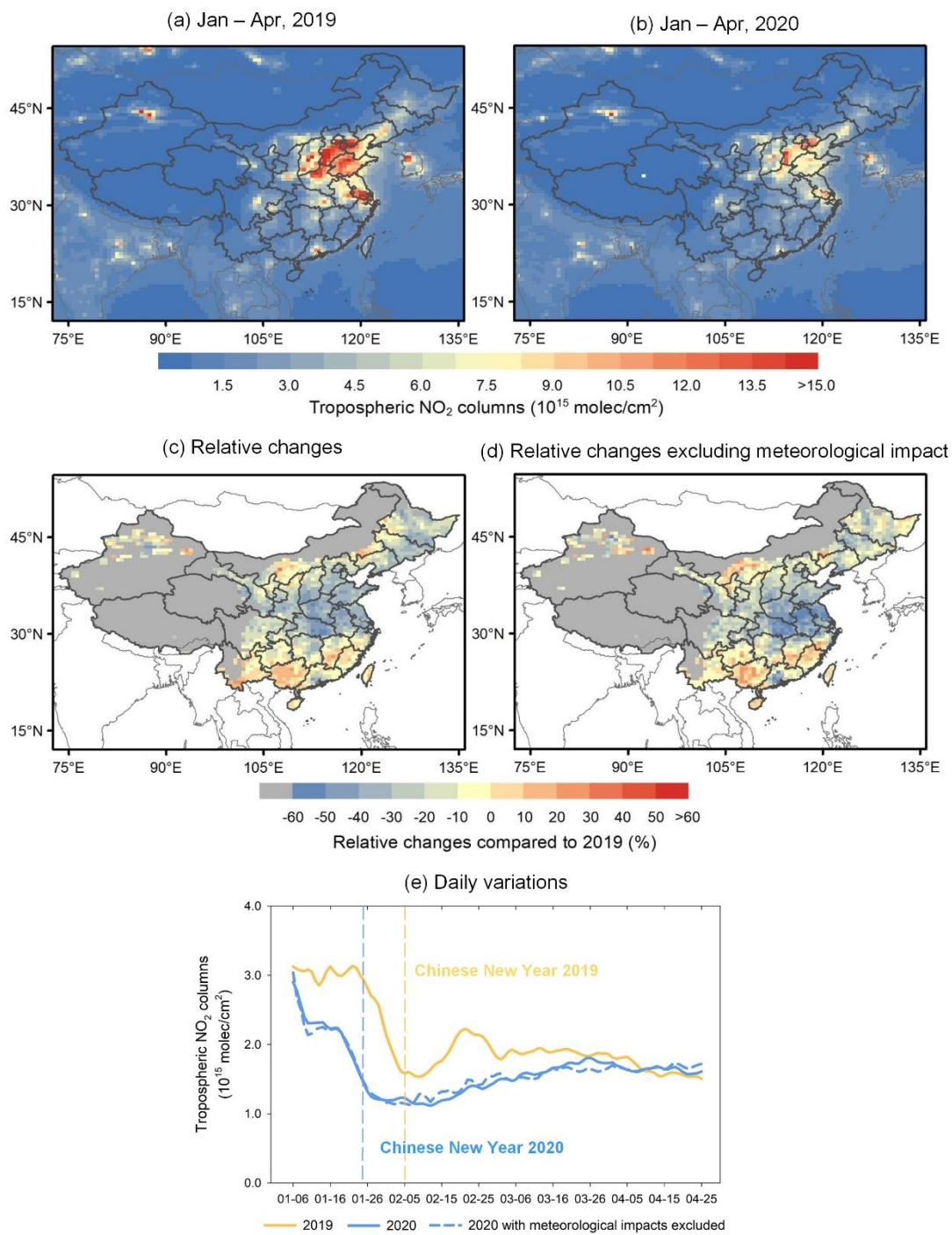

**Fig. S6. Tropospheric NO$_2$ column densities from TROPOMI over China in 2019 and 2020.** Average NO$_2$ TVCDs during Jan to Apr in (a) 2019 and (b) 2020. (c) Total relative differences between 2019 and 2020 and (d) Relative change between 2019 and 2020 excluding the influence from meteorological variations. (e) Daily variations in 10-day moving average of NO$_2$ TVCDs from TROPOMI over China.

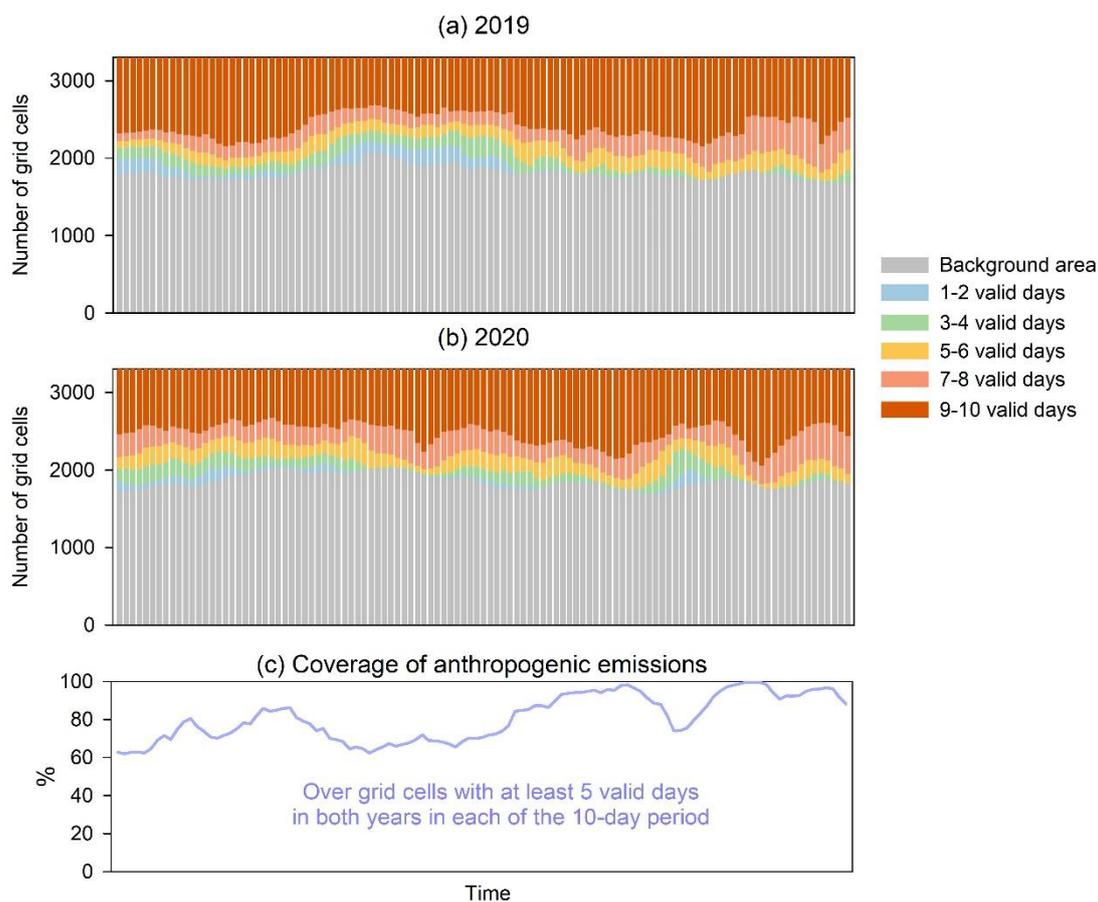

**Fig. S7. Sample size in the 10-day moving average calculation of satellite NO$_2$ TVCDs and representation of national NO$_x$ emissions.** Number of grid cells that are filtered out (i.e., $<1\times10^{15}$ molec/cm$^2$) (grey) or have certain number of valid days in (a) 2019 and (b) 2020. (c) The coverage of anthropogenic emissions on each day over grid cells with at least 5 valid days in both years in the 10-day moving average calculation.

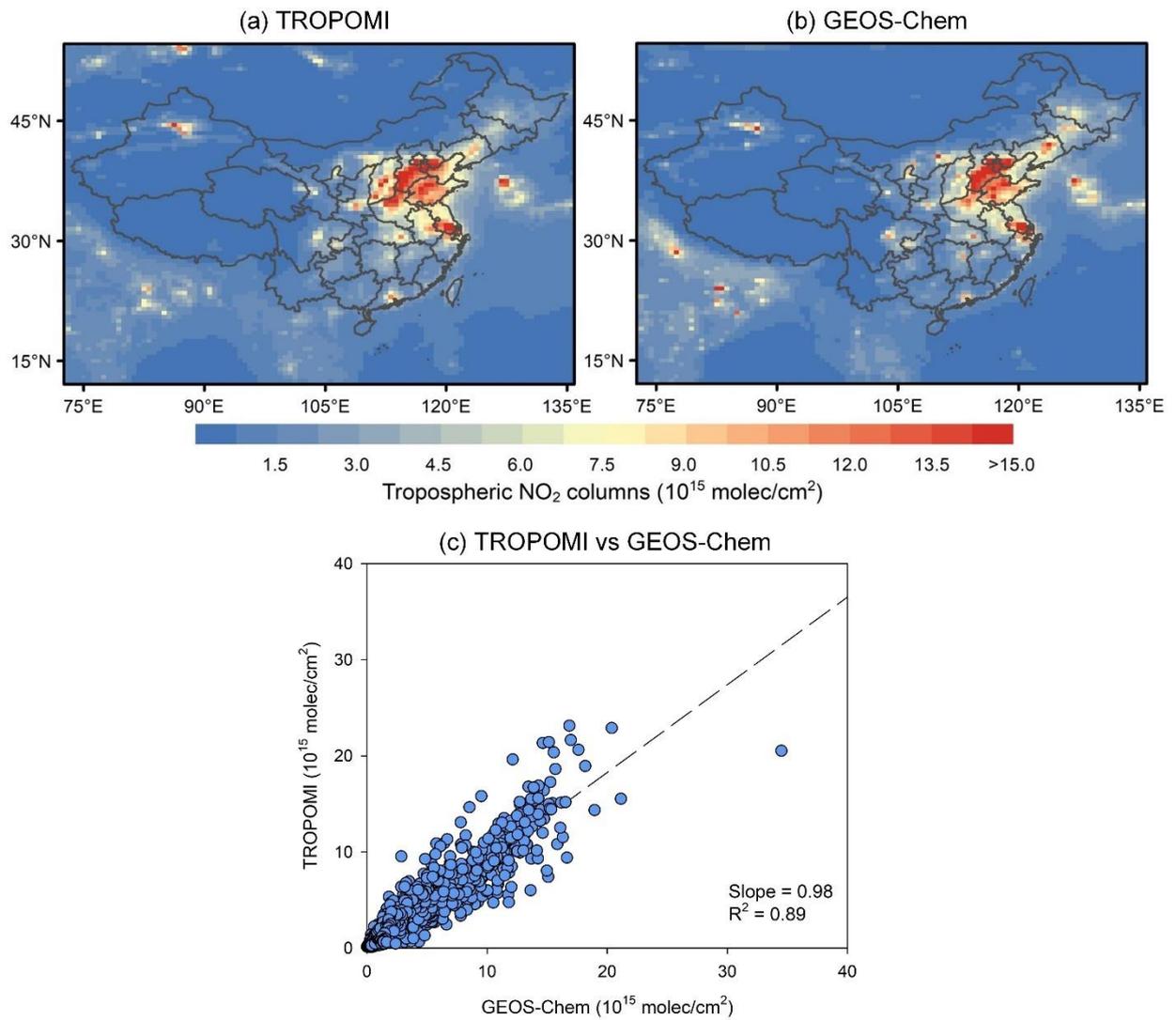

**Fig. S8. Evaluation of the GEOS-Chem baseline scenario with NO$_2$ TVCDs from TROPOMI**. Spatial distribution of coincidently sampled four-month averaged NO$_2$ TVCDs from (a) TROPOMI and (b) GEOS-Chem. (c) Scatter plot between model simulations and TROPOMI retrievals, with regression R$^2$ and slope displaced in the panel.

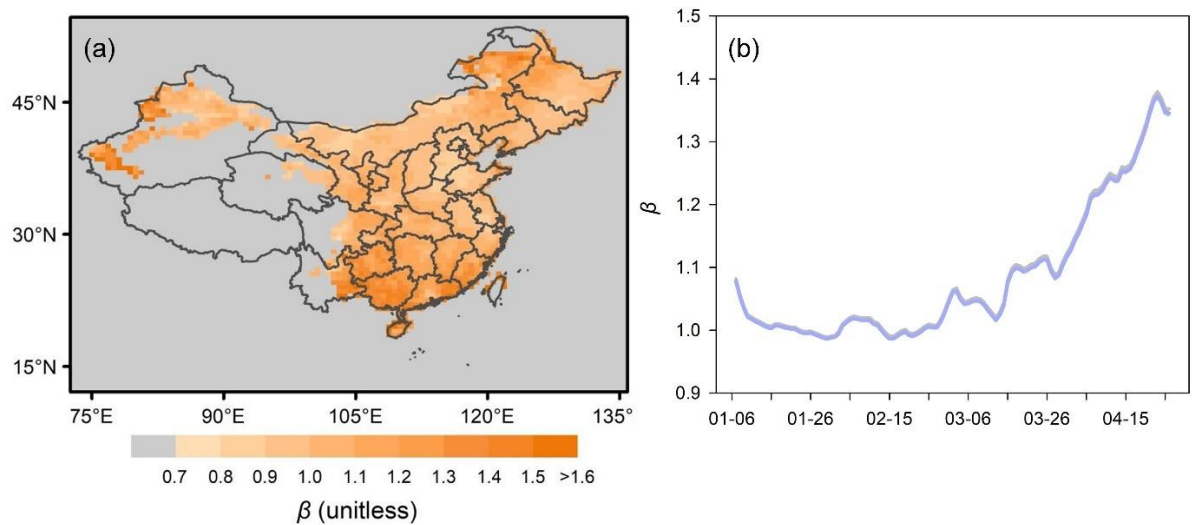

**Fig. S9.** **(a) Spatial distribution and (b) daily variations of beta values over China estimated from GEOS-Chem model.** The daily variations are also 10-day moving average.

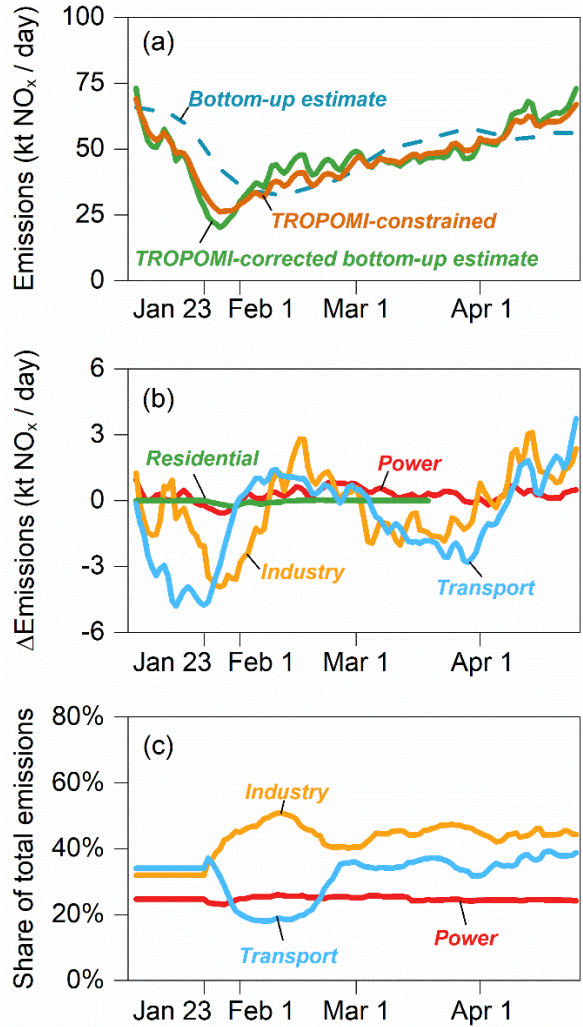

**Fig. S10. Comparison of bottom-up NO$_x$ emissions with the TROPOMI inversions in 2020.** (a) Comparison of bottom-up estimated NO$_x$ emissions, TROPOMI-constrained NO$_x$ emissions, and TROPOMI-corrected bottom-up estimate. (b) The difference in NO$_x$ emissions between the TROPOMI-constrained estimates and the bottom-up estimates over the grid cells dominated by each source sector. (c) The share of emissions from the grid cells dominated by one source sector in the total emissions of this sector. All of the data are shown in the 10-day moving average.

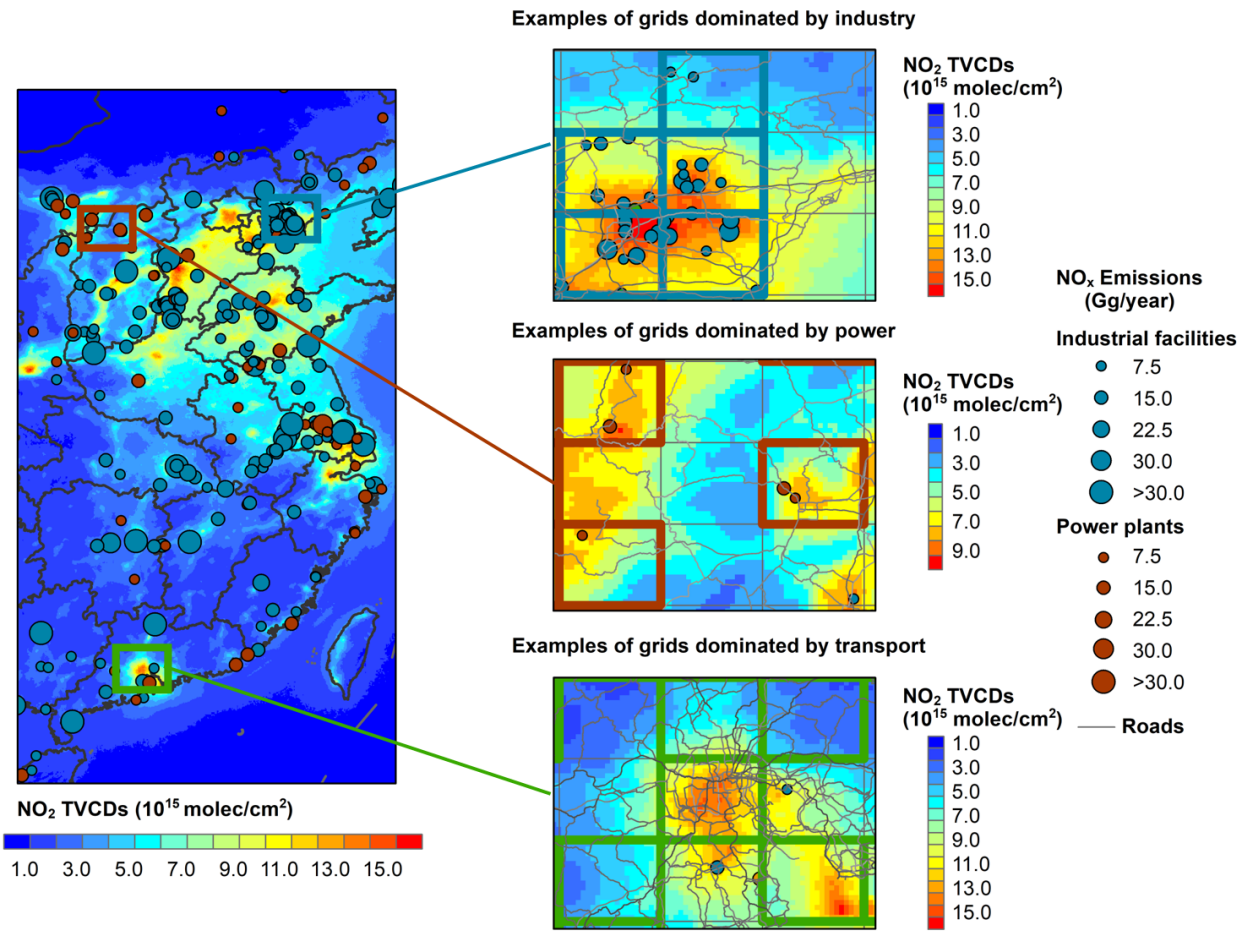

**Fig. S11. TROPOMI NO$_2$ columns over the GEOS-Chem grid cells dominated by industry, power, and transport sectors.** The dots represent the point sources of NO$_x$ emissions from industrial facilities and power plants. We also show road networks in the figures.

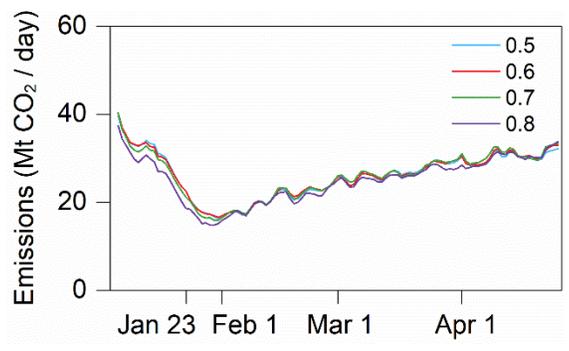

**Fig. S12. The comparison of CO$_2$ inversion emissions derived from different threshold values for the definition of the dominant emission source sector.**

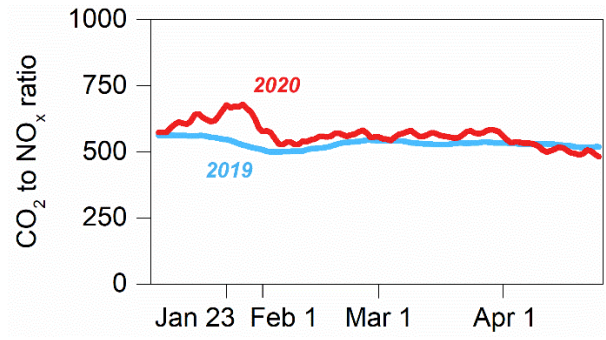

**Fig. S13. Comparison of 10-day mean $CO_2$ to $NO_x$ emission ratio between 2019 and 2020.**

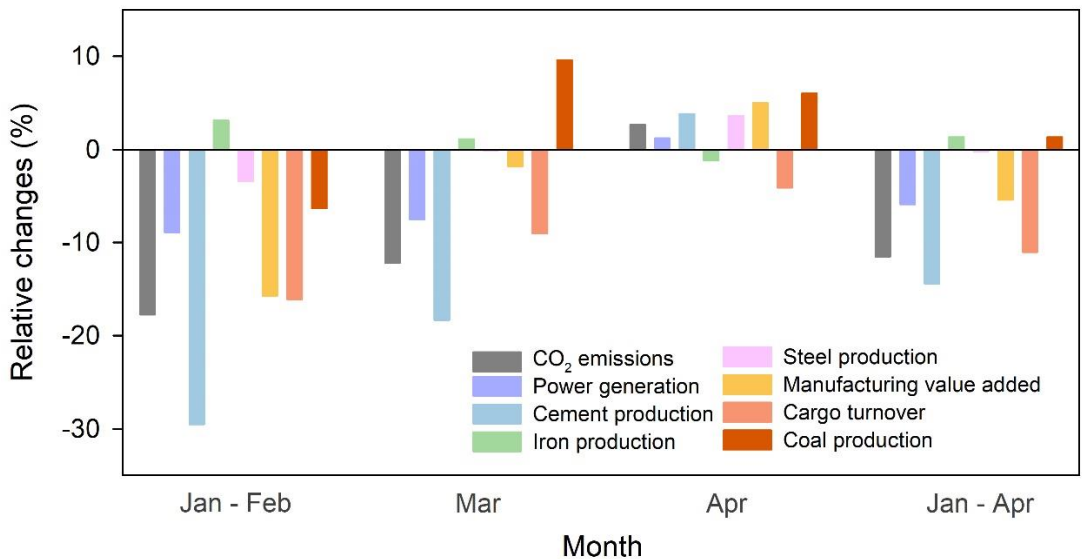

**Fig. S14. Relative changes in CO$_2$ emissions and industry activities from 2019 to 2020.** The CO$_2$ emissions are derived from the TROPOMI-constrained emissions in this study. The other data that represent the industrial activities in China are all derived from National Bureau of Statistics (http://www.stats.gov.cn/). The comparison between emissions and industrial indicators is made on monthly time scales for Jan-Feb, Mar, Apr, and Jan-Apr.

**Table S1. Daily average emissions of $NO_x$ and $CO_2$ for each month in 2019 and 2020**

|  | $NO_x$ (Mt $NO_2$) | | | $CO_2$ (Mt $CO_2$) | | |
| --- | --- | --- | --- | --- | --- | --- |
| **Month** | 2019 | 2020 | Diff | 2019 | 2020 | Diff |
| **January** | 1.6 | 1.1 | -26.9% | 854.3 | 704.3 | -17.6% |
| **February** | 1.4 | 1.1 | -23.3% | 728.3 | 598.0 | -17.9% |
| **March** | 1.8 | 1.5 | -17.0% | 948.8 | 833.6 | -12.1% |
| **April** | 1.4 | 1.5 | 5.1% | 736.2 | 755.8 | 2.7% |
| **January to April** | 6.1 | 5.2 | -15.9% | 3267.6 | 2891.6 | -11.5% |

## References

1. M. Li *et al.*, Anthropogenic emission inventories in China: a review. *National Science Review* **4**, 834-866 (2017).
2. B. Zheng *et al.*, Trends in China's anthropogenic emissions since 2010 as the consequence of clean air actions. *Atmos. Chem. Phys.* **18**, 14095-14111 (2018).
3. D. Chen *et al.*, Regional CO pollution and export in China simulated by the high-resolution nested-grid GEOS-Chem model. *Atmospheric Chemistry and Physics*, (2009).
4. Y. X. Wang, M. B. McElroy, D. J. Jacob, R. M. Yantosca, A nested grid formulation for chemical transport over Asia: Applications to CO. *Journal of Geophysical Research: Atmospheres* **109**, (2004).
5. J. Veefkind *et al.*, TROPOMI on the ESA Sentinel-5 Precursor: A GMES mission for global observations of the atmospheric composition for climate, air quality and ozone layer applications. *Remote Sensing of Environment* **120**, 70-83 (2012).
6. F. Liu *et al.*, High-resolution inventory of technologies, activities, and emissions of coal-fired power plants in China from 1990 to 2010. *Atmospheric Chemistry Physics* **15**, 13299-13317 (2015).
7. D. Tong *et al.*, Current Emissions and Future Mitigation Pathways of Coal-Fired Power Plants in China from 2010 to 2030. *Environmental Science & Technology* **52**, 12905-12914 (2018).
8. B. Zheng *et al.*, Infrastructure Shapes Differences in the Carbon Intensities of Chinese Cities. *Environmental Science & Technology* **52**, 6032-6041 (2018).
9. B. Zheng *et al.*, High-resolution mapping of vehicle emissions in China in 2008. *Atmospheric Chemistry Physics* **14**,   (2014).
10. L. Peng *et al.*, Underreported coal in statistics: A survey-based solid fuel consumption and emission inventory for the rural residential sector in China. *Applied Energy* **235**, 1169-1182 (2019).
11. B. Zheng *et al.*, Resolution dependence of uncertainties in gridded emission inventories: a case study in Hebei, China. *Atmospheric Chemistry and Physics* **17**, 921-933 (2017).
12. G. Geng *et al.*, Impact of spatial proxies on the representation of bottom-up emission inventories: A satellite-based analysis. *Atmospheric Chemistry & Physics* **17**, (2017).
13. V. Shah *et al.*, Effect of changing NOx lifetime on the seasonality and long-term trends of satellite-observed tropospheric NO2 columns over China. *Atmos. Chem. Phys.* **20**, 1483-1495 (2020).
14. M. Li *et al.*, Comparison and evaluation of anthropogenic emissions of SO2 and NOx over China. *Atmos. Chem. Phys.* **18**, 3433-3456 (2018).
15. Z. Liu *et al.*, COVID-19 causes record decline in global $CO_2$ emissions. Preprint at https://arxiv.org/abs/2004.13614 (2020).
16. Center for International Earth Science Information Network - CIESIN - Columbia University. (NASA Socioeconomic Data and Applications Center (SEDAC), Palisades, NY, 2018).

17. H. Hersbach *et al.*, The ERA5 Global Reanalysis. *Quarterly Journal of the Royal Meteorological Society*.
18. M. Crippa *et al.*, High resolution temporal profiles in the Emissions Database for Global Atmospheric Research. *Scientific Data* **7**, 121 (2020).
19. J. van Geffen *et al.*, S5P TROPOMI NO2 slant column retrieval: method, stability, uncertainties and comparisons with OMI. *Atmos. Meas. Tech.* **13**, 1315-1335 (2020).
20. D. L. Goldberg *et al.*, Enhanced Capabilities of TROPOMI NO2: Estimating NO X from North American Cities and Power Plants. *Environmental science & technology* **53**, 12594-12601 (2019).
21. D. Griffin *et al.*, High‐resolution mapping of nitrogen dioxide with TROPOMI: First results and validation over the Canadian oil sands. *Geophysical Research Letters* **46**, 1049-1060 (2019).
22. F. Liu *et al.*, Recent reduction in NO x emissions over China: synthesis of satellite observations and emission inventories. *Environmental Research Letters* **11**, 114002 (2016).
23. L. Lamsal *et al.*, Application of satellite observations for timely updates to global anthropogenic NOx emission inventories. *Geophysical Research Letters* **38**, (2011).
24. T. W. Walker *et al.*, Trans-Pacific transport of reactive nitrogen and ozone to Canada during spring. *Atmospheric Chemistry and Physics* **10**, 8353-8372 (2010).
25. E. V. Berezin *et al.*, Multiannual changes of $CO_2$ emissions in China: indirect estimates derived from satellite measurements of tropospheric $NO_2$ columns. *Atmos. Chem. Phys.* **13**, 9415-9438 (2013).
26. D. Gu, Y. Wang, R. Yin, Y. Zhang, C. Smeltzer, Inverse modelling of $NO_x$ emissions over eastern China: uncertainties due to chemical non-linearity. *Atmospheric Measurement Techniques* **9**, 5193 (2016).
27. R. Gelaro *et al.*, The modern-era retrospective analysis for research and applications, version 2 (MERRA-2). *Journal of Climate* **30**, 5419-5454 (2017).
28. J.-T. Lin, M. B. McElroy, Impacts of boundary layer mixing on pollutant vertical profiles in the lower troposphere: Implications to satellite remote sensing. *Atmospheric Environment* **44**, 1726-1739 (2010).
29. M. Li *et al.*, MIX: a mosaic Asian anthropogenic emission inventory under the international collaboration framework of the MICS-Asia and HTAP. *Atmospheric Chemistry Physics* **17**, (2017).
30. L. T. Murray, D. J. Jacob, J. A. Logan, R. C. Hudman, W. J. Koshak, Optimized regional and interannual variability of lightning in a global chemical transport model constrained by LIS/OTD satellite data. *Journal of Geophysical Research: Atmospheres* **117**, (2012).
31. R. Hudman *et al.*, Steps towards a mechanistic model of global soil nitric oxide emissions: implementation and space based-constraints. *Atmospheric Chemistry & Physics* **12**, (2012).
32. M. Bauwens *et al.*, Impact of coronavirus outbreak on $NO_2$ pollution assessed using TROPOMI and OMI observations. *Geophysical Research Letters*, e2020GL087978.

33. M. Liu *et al.*, A new TROPOMI product for tropospheric $NO_2$ columns over East Asia with explicit aerosol corrections. *Atmos. Meas. Tech. Discuss.* **2020**, 1-22 (2020).